\documentclass[aps,prl,preprint,]{revtex4}

\usepackage{graphicx}
\usepackage{hyperref}

\newcommand{\seqn}{{Schr\"odinger equation}}

\begin{document}

\title{A Stern-Gerlach Experiment in Time}
\author{John Ashmead}
\email{akmed@voicenet.com}

\date{\today}

\begin{abstract}

In non-relativistic quantum mechanics,
path integrals are normally derived from the \seqn.
This assumes the two 
formalisms are equivalent.
Since time plays a very different role in
the \seqn\ and in path integrals,
this may not be the case.

We here derive path integrals directly by
imposing two requirements:
correct 
behavior in the classical limit
and
the most complete practicable symmetry between time and space.

With these requirements,
the path integral formalism 
predicts quantum fluctuations over the time dimension
analogous to the quantum fluctuations seen over the three space dimensions.
For constant potentials 
there is no effect.
But the coupling between rapidly varying electromagnetic fields
and the quantum fluctuations in time
should be detectable.

We consider a variation on the Stern-Gerlach experiment
in which
a particle with a non-zero electric dipole moment
is sent through a rapidly varying
electric field, oriented parallel to the particle's trajectory.
The
\seqn\ predicts changes to the precession frequency
of the wave function about
the trajectory
but no physical splitting of the beam.
With the approach here,
path integrals predict the changes to the precession frequency
and in addition that the beam
will be split in velocity and time.

\end{abstract}

\maketitle

\section{Introduction}
\subsection{Problem}
In the context of non-relativistic quantum mechanics,
path integrals are normally derived directly from the \seqn\ or,
at least,
validated against it\cite{Feynman:1965d,Schulman:1981,Swanson:1992,Khandekar:1993}.
This approach makes the implicit assumption that the two 
formalisms are equivalent.
Since the views of time most naturally associated with
the \seqn\ and with path integrals are very different,
this implicit assumption is not entirely unproblematic.

\subsection{Different views of time in \seqn\ and path integrals}

To lay a foundation,
we give a brief review of a typical derivation
of path integrals from the \seqn.
We start with the time 
independent \seqn\ for one particle
\footnote{We use natural units ($\hbar=c=1$) and take the metric 
as having signature ($1,-1,-1,-1$).
Except as noted,
the summation 
convention is in effect:
repeated Greek indices are summed from 0 to 3;
repeated Roman from 1 to 3.
Overdots, i.e. $\dot {f}$,
are used to indicate differentiation 
with respect to the current path parameter
(either the proper time along the path or else time as defined in the lab frame).
Overbars, i.e. $\bar {x}_i$,
are used to indicate classical trajectories.}
\begin{equation}
i{\partial \over {\partial t}}\psi \left( {t,\vec x} \right)=-{1 \over {2m}}\nabla ^2\psi \left( {t,\vec x} \right)+V\left( {\vec x} \right)\psi \left( {t,\vec x} \right).
\label{eq.piq1.seqn2}
\end{equation}
For infinitesimal times we 
can use this to write
 $\psi(t+\Delta{}t)$ in terms of $\psi(t)$
\begin{equation}
\psi \left( {t+\Delta t,\vec x} \right)\approx\psi \left( {t,\vec x} \right)+{{i\Delta t} \over {2m}}\nabla ^2\psi \left( {t,\vec x} \right)-i\Delta tV\left( {\vec x} \right)\psi \left( {t,\vec x} \right).
\label{eq.piq1.timeop}
\end{equation}
If a function
$\psi \left( { x'} \right)$
is sufficiently smooth,
we can expand it around a point $x$
as
\begin{equation}
\psi \left( {x'} \right)\approx \psi \left( x \right)+\left( {x'-x} \right){{d\psi \left( x \right)} \over {dx}}+{1 \over 2}\left( {x'-x} \right)^2{{d^2\psi \left( x \right)} \over {dx^2}}
\label{eq.piq1.expandpsi}
\end{equation}
which lets us write
\begin{equation}
\psi \left( x \right)-{1 \over {4ia}}{{d^2\psi \left( x \right)} \over {dx^2}}\approx \sqrt {{{-ia} \over \pi }}\int {dx'\,\exp \left( {ia\left( {x'-x} \right)^2} \right)\psi \left( {x'} \right)}.
\label{eq.piq1.nabla2gauss}
\end{equation}
By taking $a$ as 
${m \over {2\Delta t}}$
and using
the identity $1+\delta\approx{}e^{\delta}$ for small $\delta$
we may write (\ref{eq.piq1.timeop}) as
\begin{equation}
\psi \left( {t+\Delta t,\vec x} \right)\approx \sqrt {{m \over {2\pi i\Delta t}}}^3\int {d\vec x'\,\exp \left( {im{{\left( {\vec x-\vec x'} \right)^2} \over {2\Delta t}}-i\Delta tV\left( {\vec x'} \right)} \right)\psi \left( {t,\vec x'} \right)}.
\label{eq.piq1.onestep}
\end{equation}
If we do this repeatedly, 
we can push 
$\psi \left( {t} \right)$
forwards in time,
one $\Delta{}t$ at a time,
to get the value of 
$\psi \left( {t'} \right)$
at arbitrary times
\begin{equation}
\psi \left( {t'',\vec x''} \right)=\int {d\vec x'\,K\left( {t'',\vec x'';t',\vec x'} \right)\psi \left( {t',\vec x'} \right)}
\end{equation}
where the kernel $K\left( {t'',\vec x'';t',\vec x'} \right)$ represents the repeated application
of (\ref{eq.piq1.onestep}).
The kernel is then given by the Trotter product formula
\begin{equation}
\mathop {\lim }\limits_{N\to \infty }\int {d\vec x_1\ldots d\vec x_{N-1}
	 {\sqrt {{m \over {2\pi i\varepsilon }}}^{3N}}}
	 \exp \left( {{{i\varepsilon } }\sum\limits_{j=0}^{N-1} {\left( {{m \over 2}\left( {{{\vec x_{j+1}-\vec x_j} \over \varepsilon }} \right)^2-V \left( {\vec x_j} \right)} \right)}} \right)
\label{eq.piq1.kernel4psi}
\end{equation}
with
\begin{equation}
\varepsilon \equiv {{t''-t'} \over N}.
\label{eq.piq1.epsilon}
\end{equation}
The summand is the discrete form
of the non-relativistic time-independent Lagrangian
\begin{equation}
{m \over 2}\left( {{{\vec x_{j+1}-\vec x_j} \over \varepsilon }} \right)^2-V\left( {\vec x_j} \right)\mathrel{\mathop{\kern0pt\longrightarrow}\limits_{\varepsilon \to 0}}{m \over 2}\vec v^2-V\left( {\vec x} \right)=L\left( {\vec x,\vec v} \right)
\label{eq.piq1.disclagrangian}
\end{equation}
so we may identify the argument of the exponential as $i$ times the
classical action $S$
\begin{equation}
S\left[ \pi \right]\equiv \int {dtL\left[ \pi \right]}
\label{eq.piq1.classicalaction}
\end{equation}
and identify the product of integrations as a sum over paths $D\left[\pi\right]$
\begin{equation}
K\left( {t'',\vec x'';t',\vec x'} \right)=\int {D\left[ \pi \right]e^{iS\left[ \pi \right]}}
\label{eq.piq1.pathintegral2}
\end{equation}
where the paths $\pi$ are defined in terms of their coordinates
at a series of discrete times.
We have used a very intuitive view of time:
we defined the wave function at one time,
then pushed it forward step by step till we arrived at the wave function
at any later time
\footnote{
QED uses a similar approach:
the canonical commutation relations are used to
define a Hamiltonian
that we then use to march the fields forward in time.
}.

But when we look at the final expression,
it is just as natural to see it as defined over time.
The paths are naturally defined as trajectories over the domain
$t'$ to $t''$ 
and the action as a functional over such trajectories.
We are free to see the final product 
from the ``block universe perspective,''
in which we see all time as existing at once
(even if we normally experience it sequentially)
\footnote{For a particularly clear review of the ``block universe,'' Nahin\cite{Nahin:1999}.}.

The question then is what happens 
if we develop path integrals by taking
(\ref{eq.piq1.pathintegral2})
as the starting point?

\subsection{Normalization}

When we replaced the $\nabla^2$ in (\ref{eq.piq1.nabla2gauss}) 
with a Gaussian integral
we chose the normalization constant
$a={m \over {2\Delta{}t}}$
using the quietly popular ``whatever-works'' methodology.
In general, a certain arbitrariness about normalization seems to be
a common feature of path integrals
\footnote{These difficulties are present in field theory as well.
As Weinberg\cite{Weinberg:1995} 
puts it ``\ldots although the path-integral formalism provides us with manifestly Lorentz-invariant rules, it does not make clear why the S-matrix calculated in this way is unitary. As far as I know, the only way to show that the path-integral formalism yields a unitary S-matrix is to use it to reconstruct the canonical formalism, in which unitarity is obvious.''}.
The one unavoidable normalization requirement is that the sum of 
the probabilities of all possibilities
be one
\footnote{Not completely unavoidable: see
Calderon's \textit{Life is a Dream}\cite{Calderon}.},
expressed in quantum mechanical terms as unitarity
\begin{equation}
\forall a,1=\sum\limits_{\left\{ b \right\}} {P\left( {b\left| a \right.} \right)}=\sum\limits_{\left\{ b \right\}} {\left| {K\left( {b;a} \right)} \right|^2}
\label{eq.piq1.largeunitarity}
\end{equation}
where $a$ represents the starting state(s) 
and ${\left\{b\right\}}$ the set of all possible outcomes.
In the case of the \seqn, (\ref{eq.piq1.largeunitarity}) is enforced at every time $t$.
Assuming a wave function is properly normalized and obeys the \seqn,
it will satisfy
\begin{equation}
{{dp \left( t \right)} \over {dt}}=0
\label{eq.piq1.constancyofrho}	
\end{equation}
where $p$ is the probability
\begin{equation}
p \left( t \right)\equiv \int {d\vec x\,\psi ^*\left( {t,\vec x} \right)\psi \left( {t,\vec x} \right)}.
\label{eq.piq1.integralofrho}
\end{equation}

We are requiring that $p(t)=1$ at all times from $t'$ to $t''$, inclusive.
Given that we only have direct
knowledge of the probabilities at the endpoints $t'$ and $t''$,
insisting that $p(t)=1$ at all times in between
is a stronger requirement than is strictly necessary.
And there are two specific problems 
with normalizing on these intermediate hypersurfaces.

The first problem is the implicit selection of a specific set of spacelike hypersurfaces
on which
to define the probability density.
Such a selection
is not manifestly invariant under all possible Lorentz transformations.
This is not in itself provably wrong
-- for one thing the \seqn\ is only supposed to be valid for non-relativistic 
quantum mechanics --
but it is troubling.

One of the troubled is Suarez,
who raised
the possibility
that if we assume standard quantum mechanics 
is correct,
with sufficient ingenuity
we could demonstrate retrotemporal causal influences.
He proposed a specific alternative to standard quantum mechanics,
relativistic nonlocality (RNL),
to avoid this difficulty (and some others)
\cite{Suarez:1997,Suarez:1997b,Suarez:1998,Suarez:1998b,Suarez:1998c,Suarez:1998d,Suarez:1998e,Suarez:2000}.
RNL
has been refuted experimentally by Stefanov, \textit{et al} \cite{Stefanov:2001}.
But the experimental refutation of RNL does not reduce the force
of Suarez's original objections; 
it merely indicates that one possible resolution of them does not work.

The second problem 
is that in the \seqn, we are overlooking the possibility of quantum jumps in time.
As we know,
there is nothing quantum mechanical systems 
enjoy more than tunneling through barriers in space.
Given that time and space are to a large extent
interchangeable, 
we might expect that
if quantum particles tunnel through barriers in space freely, 
they might ``tunnel in time'' as well.
That is to say, 
they might not be completely well-defined by their wave functions on any 
specific spacelike hypersurface.


\section{Assumptions}

Before developing our approach
we need to explicate the assumptions on which it will be based.
While this is always sound practice,
in any discussion involving the nature of time
it is imperative.
As Schulman noted
in his reply\cite{Schulman:2000b} to Casati, Chirikov, and
Zhirov's response\cite{Casati:2000} to his
``Opposite Thermodynamic Arrows of Time''\cite{Schulman:1999}:
``They and I find ourselves in a situation that is common in discussions of the `arrow of time,' namely no disagreement on technical issues and no agreement on basic assumptions.''

We will be playing a game of ``as if'' based on three assumptions:
\begin{enumerate}
 \item time is a kind of space (time/space symmetry),
 \item all time is to be seen at once (block universe perspective), and
 \item there is no fundamental direction to time (time reversal symmetry).
\end{enumerate}
Because of the importance of these points to our argument
we briefly review the evidence for them.

\subsection{Time/space symmetry}

As Minkowski\cite{Minkowski:1908} famously put it,
``Henceforth space by itself, and time by itself, 
are doomed to fade away into mere shadows, and only a kind of union of the two will preserve an independent reality.''
This approach is fundamental to special and general relativity;
for an extended review see Barbour's \textit{The End of Time}\cite{Barbour:2000}.
While the situation in quantum mechanics
is less clear,
there is nothing to refute this principle here either;
for an extended review see Stenger's \textit{Timeless Reality}\cite{Stenger:2000}.

In general, phenomena seen for space (or space and momentum)
are seen for time (or time and energy),
e.g., diffraction and interference effects and the uncertainty principle.

Diffraction and interference effects 
reinforce the view that time and space are interchangeable.
Just as we see diffraction and interference in space, we see them in time.
The first article was Moshinsky\rq{}s appropriately named ``Diffraction in time''
\cite{Moshinsky:1952}.
Experimental confirmations include
\cite{Hils:1998,Hauser:1974,Bernet:1996,Szriftgiser:1996,Garcia:2002,Godoy:2002}.
``Quantum beats''
-- self-interference in time --
have also been predicted and seen \cite{Rauch:1986,Bluhm:1996,Bluhm:1997}.

And just as we have uncertainty relations between space and momentum,
	we have uncertainty relations between time and energy.
An uncertainty principle between time and energy was proposed by
	Heisenberg (\cite{Heisenberg:1927} as translated in \cite{Heisenberg:1983}).
This played a critical role in establishing the principles of the quantum.
Einstein\rq{}s attempt to refute the uncertainty principle using his celebrated ``clock-in-a-box'' experiment
	was in turn refuted by Bohr
	using the time/energy uncertainty relation 
	(and also general relativity!: see for instance Bohr\rq{}s account in \cite{Bohr:1949}).
If there is not some kind of an uncertainty principle between time and energy,
a serious attack on the self-consistency of quantum mechanics could easily be mounted.

A particularly cogent review of the uncertainty principle
between time and energy is given by Busch\cite{Busch:2001}.
He argues that the time/energy uncertainty relationship is valid
	but does not stand on quite the same footing at the position/momentum one.
(We will be arguing this is because the standard quantum mechanics formalism
is asymmetric in its handling of time and space.
We will be proposing a more symmetric approach.)

Hilgevoord \cite{Hilgevoord:1996, Hilgevoord:1996b}
	argues that one must keep the definitions of time and space parallel
	to those for energy and momentum
	if there is to be the same kind of uncertainty principle for both.
In particular, Hilgevoord notes the need for care in distinguishing between
the time coordinate associated with the implicit space/time grid against which 
motion is being measured
and the time coordinate associated with any specific particle trajectory.
(This is an idea we take advantage of below.)

While the literature is not completely unambiguous,
we will
assume
that the time/energy uncertainty relationship stands on as firm
a basis as the space/momentum relationship,
and
therefore
that any deviation represents either an asymmetry in the experimental setup
or an asymmetry in the formalism,
the first of which should be discounted, the second eschewed.

\subsection{Block universe perspective}

Implicit in time/space symmetry is the block universe perspective.
If time is a space dimension,
then since we may consider space ``all-at-once''
we may also consider time ``all-at-once.''

We will need to invoke the block universe perspective 
when it comes time to normalize our paths.
We will be normalizing them across time,
not at each time;
this makes sense only if it makes sense to see time all at once.

The block universe perspective has dramatic experimental support
	in the form of Wheeler\rq{}s delayed choice experiment \cite{Wheeler:1978}.
This is a double-slit experiment with a twist.
In a standard double slit experiment,
	if one checks which slit the particle went through, the interference pattern is lost.
In a delayed choice experiment,
	the decision as to whether or not to 
	check which slit the particle went through 
	is made \emph{after} the particle has
	(nominally) already gone through one or both slits.
As Wheeler \cite{Wheeler:1981} put it:
``In the new \lq{}delayed-choice\rq{} version of the experiment one decides
	whether to put in the half-silvered mirror or take it out at the very last minute.
Thus one decides whether the photon \lq{}shall have come by one route,
	or by both routes\rq\ after it has \lq{}already done its travel\rq.''
The predictions of quantum mechanics have been confirmed experimentally:
	recent experiments include \cite{Kim:2000b,Walborn:2002}.
	
Wheeler has further pointed out (in his ``Great Smoky Dragon'' experiment\cite{Wheeler:1983b})
	that one may even perform a version of the delayed choice experiment
	across cosmological distances and times,
	by taking advantage of gravitational lensing effects.
	
It is in our view difficult to make sense of these results except by
taking the block universe perspective.
If we imagine the particle in question having a definite position in time,
then to reproduce the experimental results,
the particle would have to scoot ahead (in time) to see what experimental setup was waiting for it,
then dart back (in time) to take both doors or just one,
depending.
We are not saying it is beyond imagination to come up with some way to make
this plausible.
But we find it simpler to take the experimental evidence as given.

If the delayed choice experiment is insufficiently persuasive,
then we may consider the
 ``quantum eraser'' experiments,
	first proposed by Scully \cite{Scully:1982,Scully:1991b}.
We have the same experimental setup as with the delayed choice experiment,
	but now, after collecting the ``welcher weg'' or ``which path'' information
	we deliberately erase the information, while maintaining quantum coherence.
The erasure of the ``welcher weg'' information restores the original interference pattern.
Now our rather harried particle has not only to scoot ahead in time to see
 whether its path is being observed,
it has to scoot still further ahead to see if that information is being kept or discarded.
The quantum eraser has also been confirmed, see Herzog, \textit{et al} \cite{Herzog:1995}.

There are reviews of the literature for the delayed choice and quantum eraser 
experiments
in Ghose\cite{Ghose:1999} and in Auletta\cite{Auletta:2000}.

\subsection{Time reversal symmetry}

Both time/space symmetry and the block universe perspective could be brought into question
by the detection of any hard evidence for an underlying
direction to time.
But
-- provided we include $CPT$ in our definition of reversal in time
-- there does not appear to be
any evidence for such an asymmetry.
Per Rosner\cite{Rosner:2000}
``The discrete symmetries $C$ (charge inversion), $P$ (parity, or space reflection),
	and $T$ (time reversal) are preserved by strong and electromagnetic processes,
	but violated by weak decays. For a brief period of several years,
	it was thought that the products $CP$ and $T$ were preserved by all processes,
	but that belief was shattered with the discovery of $CP$ violation in
	neutral kaon decays in 1964 \cite{Christenson:1964}.
The product $CPT$ seems to be preserved,
	as is expected in local Lorentz-invariant quantum field theories.''
If we look at the motion of
a set of particles and look at the same motion after the $CPT$ operation
is applied, 
we cannot tell which was the original and which the reversed
\footnote{Since we are only looking at electrodynamics, 
we do not need the qualification about $CPT$.},
assuming the set of particles is not large enough 
that entropic considerations come into play.

We will assume for the rest of this investigation,
	that the directional character associated with time is 
	\emph{entirely} a product of entropy.
A strong entropic gradient was established by the Big Bang,
	and we have been going downhill since.
As Lebowitz\cite{Lebowitz:1995} put it
``Laboratory systems are prepared in states of low Boltzmann entropy
	by experimentalists who are themselves in low-entropy states.
Like other living beings,
	they are born in such states and maintained there by
	eating nutritious low-entropy foods,
	which in turn are produced by plants using low-entropy radiation coming from the Sun.''

Or to put it another way,
at the level of propagators and particles there is no causal direction;
	that is to be found only in collections of particles,
	i.e. beams and other impulsive things.
	``For a transmitting aerial,
	the effective temperature of the source is made much larger than that of the surroundings,
	for a receiving aerial,
	the effective temperature of the load is made much less than that of the surroundings.
There is no `one-sidedness' in nature relating to the way
	in which oscillating electrons radiate energy, as some authors imply.''
\cite{Stephenson:1978},
\footnote{This is the familiar Einstein-Ritz debate: ``While Einstein believes that one may restrict oneself to this case without essentially restricting the generality of the consideration, Ritz regards this restriction as not allowed in principle. If one accepts the latter point of view, experience requires one to regard the representation by means of the retarded potentials as the only possible one, provided one is inclined to assume that the fact of the irreversibility of radiation processes has to be present in the laws of nature. Ritz considers the restriction to the form of the retarded potentials as one of the roots of the Second Law, while Einstein believes that the irreversibility is exclusively based on reasons of probability.''\cite{Einstein:1909} as translated in Zeh's comprehensive survey\cite{Zeh:2001}.}.

If all of the directionality of time 
is given by entropy,
then if somehow a push could be administered from the opposite direction in time,
	then we might see arrows of cause and effect going both ways.
This disconcerting possibility is the subject of Schulman's
	``Opposite Thermodynamic Arrows of Time''\cite{Schulman:1999}.
He finds no irresolvable contradictions
\footnote{Of course,
when calculating
with path integrals,
it is easy to eliminate contradictions:
1) construct the set of all paths,
2) throw out all self-contradictory ones,
3) sum over the rest,
4) normalize appropriately, 
5) declare the result your answer.
And if there are no self-consistent paths,
6) declare the problem ill-posed.}.

\subsection{Transactional Interpretation}


We will rely on
Cramer's \cite{Cramer:1986,Cramer:1988b} Transactional Interpretation
of quantum mechanics.
Unlike most interpretations,
the Transactional Interpretation is manifestly consistent with our
assumptions: 
it treats time as a space dimension, 
takes the block universe perspective,
and treats the forward and backward directions in time symmetrically.

For example,
Cramer
gives as the interpretation of a particle that is emitted and
then absorbed:
``But an equally valid interpretation of the process
 is that a four-vector standing wave has been established between emitter and absorber. 
As a familiar 3-space standing wave is a superposition of waves traveling to the right and left, 
this four-vector standing wave is the superposition of advanced and retarded components. 
It has been established between the terminating boundaries of the emitter, 
which blocks passage of the advanced wave further down the time stream, 
and the absorber, which blocks passage of the retarded wave further up the time stream. 
This space-time standing wave is the transaction\ldots"

As an additional benefit, 
the Transactional Interpretation works well with path integrals.
We may take the offer as the sum $K(b;a)$ of all paths from emitter to absorber;
the acceptance as the sum $K^{\dagger}(b;a)$ of all paths from absorber back to emitter.
 
\section{Four dimensional path integrals}

\subsection{Abstract definition of the path integral}

All path integral calculations 
begin by forming the sum of all paths ${\pi \left( {b;a} \right)}$ 
from a state (or set of states) $a$ to 
a state (or set of states) $b$
\begin{equation}
K\left( {b;a} \right)=\sum\limits_{\left\{ {\pi \left( {b;a} \right)} \right\}} {e^{iS\left[ {\pi \left( {b;a} \right)} \right]}}.
\label{eq.piq1.sumofpaths}
\end{equation}
$a$ defines the set of paths consistent with whatever the preparation procedure is;
$b$ is the set consistent with a specific measurement procedure.

The probability of $b$ given $a$ is
\begin{equation}
P\left( {b\left| a \right.} \right)=K^\dagger \left( {b;a} \right)K\left( {b;a} \right).
\label{eq.probability}
\end{equation}
There is nothing in this approach to insist that the time of $a$ be less than the time
of $b$,
although in all cases we consider it will be
\footnote{No paradox or backwards-in-time travel
is implicit in selecting $t_a>t_b$.
For instance, if we were a detective attempting to retrodict the path of a quantum bullet,
we might easily wish to take the state $a$ 
as corresponding to the impact of the bullet 
and then attempt to infer the probability of all states $b$ corresponding
to its firing.}.

Our problem is to define the Lagrangian, paths,
and
normalization
appropriate to (\ref{eq.piq1.sumofpaths}).
We focus on
a single charged particle 
in an electromagnetic field.
In most cases, we will be starting at a specific point $x'$ and ending at $x''$,
both to be understood as four-vectors
($\pi \left( {b;a} \right)\equiv \pi \left( {x'';x'} \right)$).

\subsection{Lagrangian}

For any given problem
a wide range of Lagrangians 
will give the same classical trajectories.
Since we are insisting on time/space symmetry,
we will only consider Lorentz invariant Lagrangians.
For the motion of a charged particle in an electromagnetic
field, 
we may, per Goldstein\cite{Goldstein:1980},
pick any of the form
\begin{eqnarray}
L&=&-mf\left( {u^2} \right)-qu^\mu A_\mu \hfill
\label{eq.piq1.possiblelagrangian}
\end{eqnarray}
with
\begin{eqnarray}
u^\mu &\equiv& \left( {{{dt} \over {d\tau }},{{d\vec x} \over {d\tau }}} \right)=\left( {\dot {t},\dot {\vec x}} \right)\hfill\cr
 A^\mu &\equiv& \left( {\phi ,\vec A} \right)
\label{eq.piq1.definitions}
\end{eqnarray}
where $\tau$ is the proper time along the path,
provided
\begin{equation}
\left. {{{\partial f\left( y \right)} \over {\partial y}}} \right|_{y=1}={1 \over 2}.
\label{eq.piq1.condition}
\end{equation}
Since with the choice
\begin{equation}
f\left( {u^2} \right)={1 \over 2}u^2
\label{eq.piq1.quadratic}
\end{equation}
the Lagrangian is quadratic in the four-velocity $u$
-- and therefore particularly manageable --
the obvious choice is
\begin{equation}
L=-{1 \over 2}mu^\mu u_\mu -qu^\mu A_\mu 
\label{eq.piq1.lagrangian}
\end{equation}
or
\begin{equation}
L=-{1 \over 2}m\dot {t}^2+{1 \over 2}m\dot {\vec x}^2-q\dot {t}\phi +q\dot {\vec x}\cdot \vec A.
\label{eq.piq1.lagrangian4}
\end{equation}
This gives the classical equations of motion
\begin{equation}
m\ddot {t}=-q\dot {\phi} +q\dot {t}\phi _{,0}-q\dot {x}_j{\rm A}_{j,0}
\label{eq.piq1.lagrangeeqns2}
\end{equation}
\begin{equation}
m\ddot {x}_i=-q\dot {A}_i-q\dot {t}\phi _{,i}+q\dot {x}_j{\rm A}_{j,i}
\label{eq.piq1.lagrangeeqns3}
\end{equation}
While we appear to be dealing with four variables,
we can use the equations of motion to show that
\begin{equation}
{d \over {d\tau }}\left( {\dot {t}^2-\dot {\vec x}^2} \right)=0\Rightarrow \dot {t}^2-\dot {\vec x}^2=1
\label{eq.piq1.identity}
\end{equation}
and thereby eliminate $\dot {t}$ and $t$ 
in favor of $\dot {\vec x}$ and $x$.
If all we were interested in was the classical trajectories,
this would be a logical next step.
However, since it is quantum fluctuations in time
we wish to model,
the dependence on $t$ 
and the terms
\begin{equation}
-{1 \over 2}m\dot {t}^2-q\dot {t}\phi 
\label{eq.piq1.timeterms}
\end{equation}
are of the essence.
For future reference,
(\ref{eq.piq1.lagrangeeqns2},\ref{eq.piq1.lagrangeeqns3})
imply
\begin{equation}
{d \over {d\tau }}\left( {m\dot {t}} \right)={{\vec v\cdot \vec E} \over {\sqrt {1-\vec v^2}}}
\label{eq.piq1.timeforce}.
\end{equation}

While the Lagrangian (\ref{eq.piq1.lagrangian}) satisfies our requirements,
there is no guarantee it is the ``correct'' Lagrangian
from a quantum mechanical point of view.
We may think of the classical trajectory
as being like the river running through the center of a valley;
the quantum fluctuations as corresponding to the topography of
the surrounding valley.
Many different topologies of the valley are consistent with the
same course for the river.
However, as (\ref{eq.piq1.lagrangian}) 
does produce the correct classical trajectories,
is symmetric between time and space,
and is, 
of the choices given by (\ref{eq.piq1.possiblelagrangian}),
the easiest to work with,
it is the obvious one to try first.

\subsection{Paths}

When computing a path integral
we normally do the sum over all paths in space
\begin{equation}
\pi =\left( {\vec x\left( t \right)} \right)
\label{eq.piq1.pathsinspace}
\end{equation}
with $\vec x'\equiv{\vec x\left( t' \right)}$
and $\vec x''\equiv{\vec x\left( t'' \right)}$.
To achieve the greatest practicable symmetry
between time and space we have to include all paths that vary in time as well,
i.e. something more like
\begin{equation}
\pi \equiv \left( {t\left( \tau \right),\vec x\left( \tau \right)} \right).
\label{eq.piq1.paths}
\end{equation}
To achieve symmetry between $t$ and $\vec x$ we have to let the time
coordinate of a path vary just as we let the space coordinate vary.

In the normal case, when we wish to describe all the paths
from $x'$ at $t'$ to $x''$ at $t''$
(taking just one space dimension for simplicity)
we may model them in terms of their offset from a straight line
\begin{equation}
x\left( t \right)=x'+{{x''-x'} \over {t''-t'}}\left( {t-t'} \right)+\sum\limits_{n=1}^\infty {a_n\sin \left( {\omega _n\left( {t-t'} \right)} \right)}
\label{eq.piq1.xpaths}
\end{equation}
where $\omega_n$
is given by
\begin{equation}
\omega _n\equiv {{n\pi } \over {t''-t'}}
\label{eq.piq1.omegan}
\end{equation}
and the $a_n$ are real.
This guarantees that all paths that begin at $t',x'$ and end at $t'',x''$
are included:
there is a one-to-one mapping between the
set of all paths 
$\pi \left( {x'';x'} \right)$
and the set of all 
$\left\{ {a_n} \right\}$.

We would like to do the same thing for paths defined in time.
There are, however, two problems:
\begin{enumerate}
 \item What are we going to use for the time dimension? 
 If time is now a dependent variable, what should we use for the independent variable?
 \footnote{Defining time along a quantum mechanical path
is a non-trivial problem: 
for an exploration of the difficulties, 
see \cite{Sokolovski:2001} and other articles in \cite{Muga:2001b}.}
 \item And what do we mean by using time as a dependent variable?
\end{enumerate}

We address the first problem by defining the time $t$
associated with a particle trajectory
in terms of $T$,
the time in the lab frame
\footnote{Per Goldstein, we have considerable freedom 
in the choice of the parameter.
Proper time is the most obvious.
However, 
if we used proper time we 
would have $t$ and $\vec x$
parameterized in terms of $\tau$,
$t=t(\tau)$ and $\vec x=\vec {x}(\tau)$.
But $\tau$ is in turn defined 
as an integral
over $t$ and $\vec x$,
$d\tau =d\sqrt {t^2+\vec x^2}$.
This would leave us with circular definitions.}.
The idea is that the time associated with a particle trajectory 
has, like Vonnegut's Billy Pilgrim\cite{Vonnegut:1969}, become slightly ``unstuck'' 
from time as defined in the laboratory frame.
$T$ takes the role that $\tau$ has had in the Lagrangian
\footnote{From here forward, $\dot {f}$ means $df / dT$.}.
In the non-relativistic case,
$t\approx{T}$.

By using $T$ as the reference index we can now define paths that fluctuate
in time exactly as we defined paths that fluctuated in space.
We may describe them in terms of their offset from a straight line
\begin{equation}
t\left( T \right)=t'+{{t''-t'} \over {T''-T'}}\left( {T-T'} \right)+\sum\limits_{n=1}^\infty {a_n\sin \left( {\omega _n\left( {T-T'} \right)} \right)}
\label{eq.piq1.tpaths}
\end{equation}
where $\omega_n$
is given by
\begin{eqnarray}
\omega _n\equiv {{n\pi } \over {T''-T'}}
\label{eq.piq1.omegan2}
\end{eqnarray}
and the $a_n$ are real.
This will guarantee that all paths begin at $T',t'$ and end at $T'',t''$
and that all are included.
Again, there is a one-to-one mapping between the
set of all paths 
$\pi \left( {t'',t'} \right)$
and the set of all 
$\left\{ {a_n} \right\}$
\footnote{
And we must change $\tau$ to $T$
in (\ref{eq.piq1.xpaths}) giving
$x\left( T \right)=x'+{{x''-x'} \over {\Delta{}T}}\left( {T-T'} \right)+\sum\limits_{n=1}^\infty {a_n\sin \left( {\omega _nT} \right)}$.
}.

We still have the second problem,
what do we mean by using time as a dependent variable?
\begin{figure}
\begin{center}
\includegraphics[width=2in]{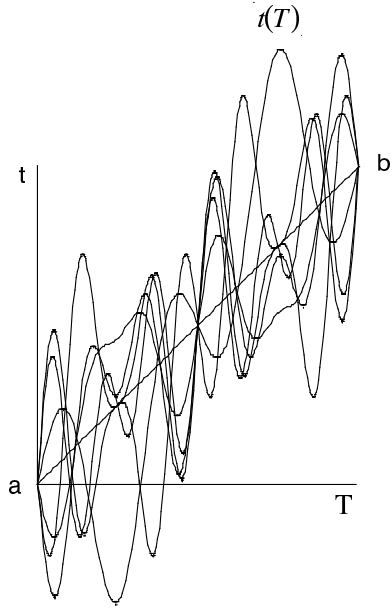}
\caption{Typical paths in time }
\label{fig:typicalpaths}
\end{center}
\end{figure}
In particular, 
there is nothing to keep the $t(T)$ in (\ref{eq.piq1.paths}) from being less than $T'$ 
or more than $T''$ at various points in its trajectory (see fig. \ref{fig:typicalpaths}).
This is unavoidable, given we are treating time as a space dimension.
Paths in space are 
allowed to zig left before zagging right
and therefore paths in time
must by the assumption of time/space symmetry have the same right.
But this means that our trajectories can 
sample the electromagnetic fields \emph{before}
the trajectory starts and \emph{after} it ends.
This does not in itself create an immediate problem for causality
-- we are insisting $T''\ge{}T'$ --
but it is perhaps a bit unnerving.

The simplest solution is to treat this as a formal device
for generating experiments;
this is legitimate if unhelpful.

A second solution is to observe that these fluctuations will be of order $\hbar$
and
therefore small.
Most paths will spend little time before $t=T'$
and little time after $t=T''$.
This helps to explain why these have not been seen
\footnote{These under- and overshoots
may show up as small violations of the optical theorem.
We are not exploring this line of attack here,
but Bennett\cite{Bennett:1987,Bennett:1987b,Bennett:1987c}
has argued that small violations of the optical theorem
may already have been seen.
To be sure, 
Bennett's conclusions are not unquestioned: 
see
Valentini\cite{Valentini:1988}.},
but does not address the point of principle.

A third solution is to note
that the starting point of a path should be regarded
as just as subject to quantum uncertainty
as anything else.
We do not know, except to within 
$\delta{}t\sim\hbar / \delta{}\omega$,
where a path starts or ends.
The variation in $t(T)$ merely reflects
our unavoidable uncertainty on this point.

This in turn leads us to an interesting if perhaps outr\'e 
view of a wave function as an object extended in time
as well as in space,
where the extension in time represents our uncertainty about the
particle's position in time,
just as the extension in space represents our uncertainty
about its position in space.
Then the usual wave function $\psi$ at a crisply defined lab time $T$
is a three-dimensional average
over the ``true'' four dimensional wave function
\footnote{We now switch to treating the lab time $T$ as an index, writing $t_T$ rather than $t(T)$.}
\begin{eqnarray}
\psi ^{\left( 3 \right)}\left( {T,\vec x\left( T \right)} \right)&=&\int\limits_{-\infty }^\infty {dt_T\psi ^{\left( 4 \right)}\left( {t_T,\vec x_T} \right)}\hfill\cr
 &=&\left\langle {\psi ^{\left( 4 \right)}} \right\rangle _{t_T}
\label{eq.piq1.threedwave}
\end{eqnarray}

The \seqn\ is then describing not the full four dimensional wave function 
but a three dimensional average of it.
Assuming for the sake of argument
that this is the case,
then we may test this approach
by looking for time-time correlations too subtle to be captured by 
the time-averaged $\psi ^{\left( 3 \right)}$.

As an aside,
perhaps the simplest -- and certainly the most Machian -- approach we could take to 
the lab time is to treat the lab coordinates, $T, \vec X$, as representing
averages over the wave function of the rest of the universe,
loosely
\begin{eqnarray}
T&\equiv& \left\langle {\tilde \psi } \right|t\left| {\tilde \psi } \right\rangle \hfill\cr
 \vec {{X}}&\equiv& \left\langle {\tilde \psi } \right|\vec x\left| {\tilde \psi } \right\rangle
\label{eq.piq1.restofuniv}
\end{eqnarray}
where $\tilde \psi$ is the rest of the wave function of the universe, 
the part complementary to the $\psi$
under examination.
Then we have no absolute time,
instead time is defined by comparing the expectation of the time
operator over a (usually) small part of the wave function of the universe 
to the expectation of the time
operator over the rest of the wave function of the universe.
The difference between $t$ and $T$
is that $t$ is the time for $\psi$,
$T$ the time for $\tilde \psi$,
$\psi$'s complement.
The assumption that this is an acceptable approximation
is essentially the 
assumption
that quantum interference terms between $\psi$
and $\tilde \psi$ may be ignored,
in other words that we may ignore quantum interference terms
between experiment and the observing apparatus\footnote{
The conceptual difficulties are compounded 
by the fact that such expectation values 
are almost always computed using the assumption
that there is a well-defined space-time over which  
the relevant integrations
may be done.
How to integrate over a space-time that is not defined
until we have done the integrations?
It is like trying to make a bed while standing on it.}.

Fortunately
we need only the assumption
that the ``lab time'' is sufficiently well-defined
to use as a background grid,
a matrix of imagined marks
created by averaging over the times associated
with an Avogadro's numbers' worth of particles,
and corresponding in function to those 
faint penciled lines and pinpricks artists 
inscribe on a canvas to mark out lines of
perspective and vanishing points,
essential for the construction,
but not themselves part of the final work.

\subsection{Normalization}

If there are quantum fluctuations in time,
then we expect that the integral of the probability density 
may vary in time.
We can define a probability density at each 
laboratory time $T$
by
\begin{equation}
\rho \left( {T,\vec x_T} \right)\equiv \int {dt_T\,\psi ^*\left( {t_T,\vec x_T} \right)\psi \left( {t_T,\vec x_T} \right)}
\label{eq.piq1.rho3d}
\end{equation}
with a probability at each time of
\begin{equation}
p\left( T \right)\equiv \int {d\vec x_T\,\rho \left( {T,\vec x_T} \right)}
\label{eq.piq1.prob}
\end{equation}
We need to allow for the possibility that
\begin{equation}
{{dp\left( T \right)} \over {dT }}\ne 0
\label{eq.piq1.probvarying}
\end{equation}
while still keeping the probability normalized in some sense
\footnote{``Keeping an open mind is a virtue
-- but as the space engineer James Oberg once said, 
not so open that your brains fall out.'' -- Sagan\cite{Sagan:1995}}.

We only actually
measure probabilities at start and finish, at $T'$ and $T''$.
Hence having
\begin{eqnarray}
1&=&\int {d\vec x_{T'}\,\rho \left( {T',\vec x_{T'}} \right)}\hfill\cr
 &=&\int {d\vec x_{T''}\,\rho \left( {T'',\vec x_{T''}} \right)}.
\label{eq.piq1.rhonorm}
\end{eqnarray}
at times $T'$ and $T''$ 
is mandatory.
But requiring 
\begin{eqnarray}
1=\int {d\vec x_T\,\rho \left( {t_T,\vec x} \right)}
\label{eq.piq1.normcond}
\end{eqnarray}
for arbitrary $T$ is not.

To force (\ref{eq.piq1.rhonorm}) 
we try normalizing the probability amplitude from $a$ to $b$
with respect to the probability amplitude from $a$ to $\left\{ b \right\}$, 
the set of all possible outcomes
\begin{equation}
\tilde {K}\left( {b;a} \right)\equiv {1 \over {\sqrt {N\left( a \right)}}}K\left( {b;a} \right)
\label{eq.piq1.kerneltwiddle}
\end{equation}
where the normalization is
\begin{equation}
N\left( a \right)\equiv \sum\limits_{\left\{ b \right\}} {K^\dagger \left( {b;a} \right)K\left( {b;a} \right)}
\label{eq.piq1.kernelnormfactor}
\end{equation}
Therefore
\begin{eqnarray}
\sum\limits_{\left\{ b \right\}} {P\left( {b\left| a \right.} \right)}&=&\sum\limits_{\left\{ b \right\}} {\tilde {K}^\dagger \left( {b;a} \right)}\tilde {K}\left( {b;a} \right)\hfill\cr
 &=&\sum\limits_{\left\{ b \right\}} {{{K^\dagger \left( {b;a} \right)} \over {\sqrt {N \left( a \right)}}}}{{K\left( {b;a} \right)} \over {\sqrt {N\left( a \right)}}}\hfill\cr
 &=&{1 \over {\sum\limits_{\left\{ b \right\}} {K^\dagger \left( {b;a} \right)}K\left( {b;a} \right)}}\sum\limits_{\left\{ b \right\}} {K^\dagger \left( {b;a} \right)}K\left( {b;a} \right)\hfill\cr
 &=&1.
\end{eqnarray}

However, this does not quite work.
For instance, if we replace sums over $b$ with integrals over $\vec x''$,
and take the usual free kernel as a test case
\begin{equation}
\sqrt {{m \over {2\pi i{(T''-T')}}}}^3e^{im{{\left( {\vec x''-\vec x'} \right)^2} \over {2{(T''-T')}}}}
\label{eq.piq1.freekernel2}
\end{equation}
we get
\begin{eqnarray}
N&=&\left( {{m \over {2\pi i{(T''-T')}}}} \right)^3\int {d\vec x''\,}e^{-im{{\left( {\vec x''-\vec x'} \right)^2} \over {2{(T''-T')}}}}e^{im{{\left( {\vec x''-\vec x'} \right)^2} \over {2{(T''-T')}}}}\hfill\cr
 &=&\left( {{m \over {2\pi i{(T''-T')}}}} \right)^3\int {d\vec x''\,}\hfill\cr
 &=&\infty.
\label{eq.piq1.quickinfinity}
\end{eqnarray}
This is an unfortunate side effect of dealing with what is really a distribution.

Since the kernel is a distribution, we may hope to get control
by selecting appropriate test functions, e.g. the Gaussians
\footnote{Kaiser\cite{Kaiser:1994}
discusses how an arbitrary wave form may be analyzed
as a sum of Gaussian wavelets.}.
\begin{equation}
\varphi_a \left( {T',\vec x'} \right)\equiv {1 \over {\left( {\pi \sigma ^2} \right)^{3/4}}}\exp \left( {-{{\left( {\vec x'-\vec x_a} \right)^2} \over {2\sigma ^2}}+i\vec k_a\cdot \left( {\vec x'-\vec x_a} \right)} \right).
\label{eq.piq1.phia}
\end{equation}
These are centered at $\vec x_a$, have momentum $\vec k_a$,
and are normalized to one at $T=T'$.

We can now compute the normalization for a specific $\varphi_a$
\begin{equation}
N_a=\int {d\vec x''\,}\varphi _a^*\left( {T'',\vec x''} \right)\varphi _a\left( {T'',\vec x''} \right)
\label{eq.piq1.normtestfunction}
\end{equation}
\begin{equation}
\varphi _a\left( {T'',\vec x''} \right)\equiv \int {d\vec x'\,}K\left( {T'',\vec x'';T',\vec x'} \right)\varphi _a\left( {T',\vec x'} \right)
\label{eq.piq1.psiofkernel}
\end{equation}
We will assume we have discovered 
the kernel is proportional
to
\begin{equation}
\sqrt {{m \over {T''-T'}}}^3e^{i{m \over 2}{{\left( {\vec x''-\vec x'} \right)^2} \over {T''-T'}}}
\label{eq.piq1.kernelproportional}
\end{equation}
(the semi-classical approximation will give this, for instance)
so
\begin{equation}
\tilde {K}\left( {x'';x'} \right)={1 \over {\sqrt {N_{a}}}}\sqrt {{m \over {T''-T'}}}^3e^{i{m \over 2}{{\left( {\vec x''-\vec x'} \right)^2} \over {T''-T'}}}.
\label{eq.piq1.kerneltwiddle3}
\end{equation}
By (\ref{eq.piq1.psiofkernel})
\begin{eqnarray}
\varphi _a\left( {T'',\vec x''} \right)&=&{1 \over {\sqrt {N_{a}}}}{1 \over {\left( {\pi \sigma ^2} \right)^{3/4}}}\sqrt {{{2\pi i} \over {f\left( {T''-T'} \right)}}}^3\hfill\cr
 &&\times \exp \left( {-{{\left( {\vec x''-\vec x'} \right)^2} \over {2\sigma ^2f\left( {T''-T'} \right)}}+i{{\vec k_a\cdot \left( {\vec x''-\vec x'} \right)} \over {f\left( {T''-T'} \right)}}-i{{\vec k_a^2\left( {T''-T'} \right)} \over {2mf\left( {T''-T'} \right)}}} \right)
\label{eq.piq1.psifuture}
\end{eqnarray}
with
\begin{equation}
f\left( T \right)\equiv 1+i{T \over {m\sigma ^2}}
\label{eq.piq1.fdefn}
\end{equation}
This gives for the probability density at $T''$
\begin{eqnarray}
\rho \left( {T'',\vec x''} \right)={1 \over {N_a}}{{\left( {2\pi } \right)^3} \over {\sqrt {\pi \sigma ^2\left| {f\left( {T''-T'} \right)} \right|^2}^3}}\exp \left( {-{{\left( {\vec x''-\left( {\vec x_a+{{\vec k_a} \over m}\left( {T''-T'} \right)} \right)} \right)^2} \over {\sigma ^2\left| {f\left( {T''-T'} \right)} \right|^2}}} \right).
\label{eq.piq1.probdensity}
\end{eqnarray}
The requirement that this be normalized to one gives
\begin{equation}
N_{a}=\left( {2\pi } \right)^3.
\label{eq.piq1.normfactor}
\end{equation}
The normalized kernel is
\begin{eqnarray}
\tilde {K}\left( {x'';x'} \right)&=&{1 \over {\sqrt {2\pi }^3}}K\left( {x'';x'} \right)\hfill\cr
 &=&{1 \over {\sqrt {2\pi }^3}}\sqrt {{m \over {T''-T'}}}^3e^{i{m \over 2}{{\left( {\vec x''-\vec x'} \right)^2} \over {T''-T'}}}.
\label{eq.piq1.kernelrewrite}
\end{eqnarray}
This is the standard free kernel up to an inessential factor of
${1 \over {\sqrt i^3}}$.

In principle,
this approach to normalization is clumsier than
the standard:
we have to normalize for each specific starting wave function $\varphi(x')$.
In other words we have
\begin{equation}
K\left( {x'';x'} \right)\to {\tilde{ K}}\left[ {\varphi} \right]\left( {x'';x'} \right).
\label{eq.piq1.ktokphia}
\end{equation}
The expression for the normalization constant
for a given $\varphi$ is
\begin{equation}
N\left[ {\varphi } \right]=\int {d\vec x''\,\left| {\int {d\vec x'\,\left( {K\left( {x'';x'} \right)\varphi \left( {x'} \right)} \right)}} \right|^2}.
\label{eq.piq1.normphia}
\end{equation}
This gives us a formal method to 
determine the normalization constant
for path integrals,
letting us avoid the \textit{ad hoc} method
employed to get (\ref{eq.piq1.nabla2gauss}).

We will be making only a limited use of 
this normalization procedure;
we are primarily interested in establishing the self-consistency 
of the approach we are taking here.
But the basic idea 
-- normalize the amplitude to get to a specific outcome
by the sum of the amplitudes to get to any outcome --
is sufficiently simple 
that a number of variations on the theme are possible.
For instance, 
the initial wave function could be defined with 
respect to an observer $a$ but
then measured by an observer $b$ moving relativistically
with respect to $a$.
(It can be useful to keep in mind that $K\left(b;a\right)$
represents correlation not causality.)

\subsection{Four dimensional kernel}

The path integral
for a charged particle of mass $m$ and charge $q$
going from point $x'$ to $x''$
in the presence of fields $A_\mu(x)$
is therefore
\begin{equation}
K\left( {x'';x'} \right)=\int {D\left[ \pi \right]e^{-i\int\limits_{T'}^{T''} {dT \,{m \over 2}}u^\mu u_\mu +qu^\mu A_\mu }}
\label{eq.piq1.pathintegral3}
\end{equation}
using 
$u=\left( {u_0,\vec u} \right)$
for the four velocity 
(We will usually suppress explicit notation
of the dependence of $K$ on $T'$ and $T''$.):
\begin{eqnarray}
K\left( {x'';x'} \right)&\equiv& \mathop {\lim }\limits_{N\to \infty }\int {\prod\limits_{i=1}^{N-1} {dt_id\vec x_i}}\exp \left( {-i\varepsilon \sum\limits_{j=1}^N {\left( {mu_j^2+qu_j^\mu A_\mu \left( {x_j} \right)} \right)}} \right)\hfill\cr
 x_j&\equiv& \left( {{{t_{j+1}+t_j} \over 2},{{\vec x_{j+1}+\vec x_j} \over 2}} \right)\hfill\cr
 u_j^\mu &\equiv& \left( {{{t_{j+1}-t_j} \over \varepsilon },{{x_{j+1}-x_j} \over \varepsilon },{{y_{j+1}-y_j} \over \varepsilon },{{z_{j+1}-z_j} \over \varepsilon }} \right)\hfill\cr
 u_j^2&=&\left( {u_j^0} \right)^2-\vec u_j^2\hfill\cr
 \varepsilon &\equiv& {{T''-T'} \over N}\hfill\cr
 \left( {t_0,\vec x_0} \right)&=&\left( {t',\vec x'} \right)\hfill\cr
 \left( {t_N,\vec x_N} \right)&=&\left( {t'',\vec x''} \right)
\label{eq.piq1.pathintegral}
\end{eqnarray}

The kernel (\ref{eq.piq1.pathintegral})
satisfies our two basic requirements:
it is symmetric in time and space
and
it reproduces the classical trajectories
(as shown below).

\subsection{Comparison to three dimensional kernel}

We compare (\ref{eq.piq1.pathintegral}) to the standard three dimensional kernel
\begin{eqnarray}
K\left( {x'';x'} \right)&\equiv& \mathop {\lim }\limits_{N\to \infty }\int {\prod\limits_{i=1}^{N-1} {d\vec x_i}}\exp \left( {i\varepsilon \sum\limits_{j=1}^N {\left( {{{m\vec v_j^2} \over 2}-q\phi \left( {t_j,\vec x_j} \right)+qv_iA_i\left( {t_j,\vec x_j} \right)} \right)}} \right)\hfill\cr
 \vec v_j&\equiv& \left( {{{x_{j+1}-x_j} \over \varepsilon },{{y_{j+1}-y_j} \over \varepsilon },{{z_{j+1}-z_j} \over \varepsilon }} \right)\hfill\cr
 \varepsilon &\equiv& {{t''-t'} \over N}.
\label{eq.piq1.pathintegral3d}
\end{eqnarray}
From
(\ref{eq.piq1.pathintegral})
we can get (\ref{eq.piq1.pathintegral3d})
by
\begin{enumerate}
 \item replacing
$u_j^0A_0\to \phi$
 and 
$\vec u_j\cdot \vec A\to \vec {{v}_{j}}\cdot \vec A$,
 i.e. taking the non-relativistic limit, and
 \item eliminating the
$\int {\prod\limits_{i=1}^{N-1} {dt_i}}\exp \left( {-i\varepsilon {m \over 2}\sum\limits_{j=1}^{N-1} {\left( {u_j^0} \right)^2}} \right)\label{eq.piq1.timefactor}$,
 i.e. getting rid of the integrals over $dt$, the quantum fluctuations in time.
\end{enumerate}

The first change merely indicates that (\ref{eq.piq1.pathintegral}) is a possible 
extension to the relativistic regime
of (\ref{eq.piq1.pathintegral3d});
the second is the interesting one.
In general, 
as we will see below,
normalization keeps the $dt$ integrations
from having any effect
on the resulting kernel
unless 
the potentials mix the time and space coordinates
\footnote{From the point of view of the three-dimensional kernel,
these quantum fluctuations in time
look like a separate quantum system.
They could be handled by the method of influence 
functionals\cite{Feynman:1965d,Khandekar:1993}.}.

While the inclusion of quantum fluctuations in time 
is a novelty,
it is difficult to see how one could get a correct relativistic generalization of
the \seqn\ 
without them.
Consider a frame $A$ in which they are not present.
Only $d{\vec x}_A$ integrations are used.
Now consider a frame $B$ going by the first at, say, $1/2c$
in the $x$ direction.
In the $B$ frame,
the integrals over $dx_A$ will look like a combination of integrals
in $dx_B$ and $dt_B$,
in other words to $B$, 
they will look as if the integrals include integrals over 
quantum fluctuations in time.

\section{Semi-classical approximation}

We may define the semi-classical approximation
as the kernel which results
when we expand the Lagrangian around
the classical trajectory.
We define
\begin{eqnarray}
t&=&\bar {t}+\delta t\hfill\cr
 x_i&=&\bar {x}_i+\delta x_i
\label{eq.piq1.delta}
\end{eqnarray}
where ${\bar {t}}(T)$ and $\bar {x}_i(T)$ 
represent the classical trajectory.
We keep the first three terms of the expansion in $\delta{}x$.
This is of course exact for potentials quadratic 
in $\delta{}x$,
is a good first approximation in many other cases,
and in general can help in understanding the qualitative characteristics 
of the behavior.

We start with the Lagrangian (\ref{eq.piq1.lagrangian})
and write out the $0^{th}$, $1^{st}$, and $2^{nd}$ terms,
using integration by paths to eliminate terms linear in $\delta u$
\begin{eqnarray}
L&\approx& -{1 \over 2}m\bar {u}_\mu \bar {u}^\mu -q\bar {u}_\mu A^\mu \left( {\bar {x}} \right)\hfill\cr
 &&+\left( {m\dot {\bar {u}}_\mu +q\dot {A}_\mu \left( {\bar {x}} \right)-q\bar {u}^\kappa A_{\kappa ,\mu }\left( {\bar {x}} \right)} \right)\delta x^\mu \hfill\cr
 &&-{1 \over 2}m\delta u_\mu \delta u^\mu -q\left( {A_{\mu ,\nu }\left( {\bar {x}} \right)+A_{\mu ,\nu }\left( {\bar {x}} \right)} \right)\delta u^\mu \delta x^\nu -{1 \over 2}q\bar {u}^\kappa A_{\kappa ,\mu \nu }\left( {\bar {x}} \right)\delta x^\mu \delta x^v.
\label{eq.piq1.genl12}
\end{eqnarray}
The first line integrated over $dT$ gives the classical
action ${\bar {S}}$.
The coefficient of $\delta x$ in the second line
is the classical equation of motion. It is identically zero per 
(\ref{eq.piq1.lagrangeeqns2},\ref{eq.piq1.lagrangeeqns3}).
We know that only the antisymmetric
part of $A$, $A_{\mu ,\nu }\left( {\bar {x}} \right)-A_{\nu ,\mu }\left( {\bar {x}} \right)$,
has physical meaning.
Therefore we eliminate the 
${A_{\mu ,\nu }\left( {\bar {x}} \right)+A_{\mu ,\nu }\left( {\bar {x}} \right)}$
term with a gauge transformation 
$A_\mu \to A_\mu +\lambda _{,\mu }$
\label{eq.piq1.gauge2}
such that
\begin{eqnarray}
\partial ^\mu \partial _\mu \lambda =-{1 \over 2}\left( {A_{\mu ,\nu }\left( {\bar {x}} \right)+A_{\mu ,\nu }\left( {\bar {x}} \right)} \right)
\label{eq.piq1.gauge3}
\end{eqnarray}
reducing the Lagrangian to
\begin{eqnarray}
-{1 \over 2}m\bar {u}_\mu \bar {u}^\mu -q\bar {u}_\mu A^\mu \left( {\bar {x}} \right)-{1 \over 2}m\delta u_\mu \delta u^\mu -{1 \over 2}q\bar {u}^\kappa A_{\kappa ,\mu \nu }\left( {\bar {x}} \right)\delta x^\mu \delta x^v
\label{eq.piq1.genl13}
\end{eqnarray}

We break out the time and space parts
\begin{eqnarray}
L&=&-{1 \over 2}m\dot {\bar {t}}^2+{1 \over 2}m\dot {\bar {x}}_i\dot {\bar {x}}_i-q\dot {\bar {t}}\phi \left( {\bar {x}} \right)+q\dot {\bar {x}}_kA_k\left( {\bar {x}} \right)\hfill\cr
 &&-{1 \over 2}m\delta \dot {t}^2-{1 \over 2}\left( {q\dot {\bar {t}}\phi _{,00}\left( {\bar {x}} \right)-q\dot {\bar {x}}_kA_{k,00}\left( {\bar {x}} \right)} \right)\delta t^2\hfill\cr
 &&+\left( {q\dot {\bar {t}}\phi _{,0i}\left( {\bar {x}} \right)-q\dot {\bar {x}}_kA_{k,0i}\left( {\bar {x}} \right)} \right)\delta t\delta x_i\hfill\cr
 &&+{1 \over 2}m\delta \dot {x}_i\delta \dot {x}_i-{1 \over 2}\left( {q\dot {\bar {t}}\phi _{,ij}\left( {\bar {x}} \right)-q\dot {\bar {x}}_kA_{k,ij}\left( {\bar {x}} \right)} \right)\delta x_i\delta x_j.
\label{eq.piq1.genl14}
\end{eqnarray}
The critical term is
${q\dot {\bar {t}}\phi _{,0i}\left( {\bar {x}} \right)-q\dot {\bar {x}}_kA_{k,0i}\left( {\bar {x}} \right)}$.
If this is zero,
then we can separate the problem into its time and space parts.
The effects of the time part will then drop out during normalization,
and we will be left with just the last line,
the usual space-space path integral.

As usual,
we can get an explicit formula for the kernel
in terms of the
action for the corresponding classical problem --
at the expense of some slightly formal manipulations.

We start by Wick rotating,
replacing $t$
by $ix_4$
(so sums over the corresponding indices will run from 1 to 4).
This gives
\begin{equation}
K\left( {x'';x'} \right)\equiv i^{N-1}\mathop {\lim }\limits_{N\to \infty }\int {\prod\limits_{i=1}^{N-1} {dx_jdy_jdz_jd{x_4}_j}}\exp \left( {i\varepsilon \sum\limits_{j=1}^{N-1} {\left( {mu_j^2+qu_j^\mu A_\mu \left( {x_j} \right)} \right)}} \right)
\label{eq.piq1.pathreal}
\end{equation}
Since the Lagrangian is quadratic in the integration variables,
the integrals telescope.
Per \cite{Schulman:1981,Khandekar:1993}, these give
\begin{equation}
K\left( {x'';x'} \right)= i^{N-1}{{A^{4N}}}\sqrt {\left| {{{\partial ^2{\bar {S}}\left( {x'';x'} \right)} \over {\partial x''_\mu \partial x'_\nu}}} \right|}e^{i{\bar {S}}\left( {x'';x'} \right)}
\label{eq.piq1.propagator}.
\end{equation}
The ${A^{4N}}$ is the result of the fact that we did not include the per-step normalization factor,
${1 \over A}=\sqrt {{m \over {2\pi i\epsilon }}}$
in each integration,
as we are normalizing at the end of the calculation.
${{{\partial ^2{\bar {S}}\left( {x'';x'} \right)} \over {\partial x''\partial x'}}}$
is the van Vleck Pauli determinant of ${\bar {S}}$
\begin{equation}
\left| {\matrix{{{{\partial ^2\bar {S}} \over {\partial x''\partial x'}}}&{{{\partial ^2\bar {S}} \over {\partial x''\partial y'}}}&{{{\partial ^2\bar {S}} \over {\partial x''\partial z'}}}&{{{\partial ^2\bar {S}} \over {\partial x''\partial x'_4}}}\cr
{{{\partial ^2\bar {S}} \over {\partial y''\partial x'}}}&{{{\partial ^2\bar {S}} \over {\partial y''\partial y'}}}&{{{\partial ^2\bar {S}} \over {\partial y''\partial z'}}}&{{{\partial ^2\bar {S}} \over {\partial y''\partial x'_4}}}\cr
{{{\partial ^2\bar {S}} \over {\partial z''\partial x'}}}&{{{\partial ^2\bar {S}} \over {\partial z''\partial y'}}}&{{{\partial ^2\bar {S}} \over {\partial z''\partial z'}}}&{{{\partial ^2\bar {S}} \over {\partial z''\partial x'_4}}}\cr
{{{\partial ^2\bar {S}} \over {\partial x''_4\partial x'}}}&{{{\partial ^2\bar {S}} \over {\partial x''_4\partial y'}}}&{{{\partial ^2\bar {S}} \over {\partial x''_4\partial z'}}}&{{{\partial ^2\bar {S}} \over {\partial x''_4\partial x'_4}}}\cr
}} \right|.
\label{eq.piq1.vanvleck4}
\end{equation}

Now we undo the Wick rotation, replacing $x_4$ by $-it$.
The only effect is to replace the van Vleck Pauli determinant
by its negative
\begin{equation}
K\left( {x'';x'} \right)=-{i^{N-1}}{{A^{4N}}}\sqrt {\left| {{{\partial ^2{\bar {S}}\left( {x'';x'} \right)} \over {\partial x''_\mu \partial x'_\nu}}} \right|}e^{i{\bar {S}}\left( {x'';x'} \right)}.
\label{eq.piq1.propagator2}
\end{equation}
This is the unnormalized kernel,
with the normalization to be computed
using (\ref{eq.piq1.normphia}).
At that time the factor of $-i^{N-1}A^{4N}$, being independent of $x'$ and $x''$,
will cancel out.
Knowing this factor is doomed in any case we take as our final result
\begin{equation}
K\left( {x'';x'} \right)=\sqrt {\left| {{{\partial ^2{\bar {S}}\left( {x'';x'} \right)} \over {\partial x''_\mu \partial x'_\nu}}} \right|}e^{i{\bar {S}}\left( {x'';x'} \right)}
\label{eq.piq1.propagator3}.
\end{equation}
There are three differences between this and the standard result
\begin{enumerate}
 \item The van Vleck Pauli determinant is four by four rather than three by three,
 as it includes time-time and time-space terms.
 \item The classical action $\bar {S}$ is in general different,
 even though the classical trajectories are the same.
 The differences represent in general the difference
 between a relativistic and a non-relativistic approach.
 \item The major difference is that the resulting kernel
 is used in a different way: 
 it is applied to four rather than three dimensional wave functions,
 opening up possibilities for interference in time
 not present in the three dimensional case.
 
 We recall we are taking the standard three dimensional wave functions
 as being the average in time of the four dimensional wave functions,
 with the time dimension being understood as representing our
 uncertainty as to \emph{when} the particle is located,
 just as the first three dimensions of a wave function may be taken
 as representing our uncertainty as to \emph{where} it is located.
 
\end{enumerate}

\subsection{Free kernel} 

In the case of the free particle
(\ref{eq.piq1.propagator3})
is exact.
Since the free action is
\begin{equation}
-{m \over 2}{{\left( {t''-t'} \right)^2-\left( {\vec x''-\vec x'} \right)^2} \over {\Delta T}}
\label{eq.piq1.freeaction}
\end{equation}
the
free kernel is
\begin{eqnarray}
K\left( {x'';x'} \right)={1 \over {N_a^{{\raise3pt\hbox{$1$} \!\mathord{\left/ {\vphantom {1 2}}\right.\kern-\nulldelimiterspace}\!\lower3pt\hbox{$2$}}}}}\left( {{m \over {T''-T'}}} \right)^2e^{-i{m \over 2}{{\left( {t''-t'} \right)^2} \over {T''-T'}}+i{m \over 2}{{\left( {\vec x''-\vec x'} \right)^2} \over {T''-T'}}}
\label{eq.piq1.freekernel}
\end{eqnarray}
which we may factor into time and space pieces as
\begin{eqnarray}
K^{\left( t \right)}\left( {t'';t'} \right)&=&{1 \over {N_a^{{\raise3pt\hbox{$1$} \!\mathord{\left/ {\vphantom {1 8}}\right.\kern-\nulldelimiterspace}\!\lower3pt\hbox{$8$}}}}}\sqrt {-{m \over {i\left( {T''-T'} \right)}}}e^{-i{m \over 2}{{\left( {t''-t'} \right)^2} \over {T''-T'}}}\hfill\cr
  K^{\left( {\vec x} \right)}\left( {\vec x'';\vec x'} \right)&=&{1 \over {N_a^{{\raise3pt\hbox{$3$} \!\mathord{\left/ {\vphantom {3 8}}\right.\kern-\nulldelimiterspace}\!\lower3pt\hbox{$8$}}}}}\sqrt {{m \over {i\left( {T''-T'} \right)}}}^3e^{+i{m \over 2}{{\left( {\vec x''-\vec x'} \right)^2} \over {T''-T'}}}
\label{eq.piq1.scaledfree2}
\end{eqnarray}
where we have 
chosen
the individual and overall phases
so that 
we have the useful symmetry property
\begin{eqnarray}
K_T^{\left( t \right)}\left( {t'';t'} \right)&=&K_{-T}^{\left( x \right)}\left( {t'';t'} \right)\hfill\cr
 K_T^{\left( x \right)}\left( {x'';x'} \right)&=&K_{-T}^{\left( t \right)}\left( {x'';x'} \right).
\label{eq.piq1.kernelsym}
\end{eqnarray}

If preparation and measurement are relative to the same reference frame
--the usual case --
then $K^{(t)}$ will cancel out during
normalization.
If $t\approx T$
we will be left with a constant
factor times
the usual free kernel in three dimensions
\begin{equation}
K^{\left( 3 \right)}\left( {t'',\vec x'';t',\vec x'} \right)=\sqrt {{m \over {2\pi i\left( {t''-t'} \right)}}}^3e^{i{m \over 2}{{\left( {\vec x''-\vec x'} \right)^2} \over {t''-t'}}}.
\label{eq.piq1.freekernelusual}
\end{equation}

\subsection{Free wave function}

We still need to normalize the kernel appropriately.
Per discussion above,
this can in general
only be done relative to a
specific set of test functions.
We select for $\varphi _a$
the test functions of (\ref{eq.piq1.phia})
appropriately extended to include the time dimension
\begin{eqnarray}
\varphi _a\left( {t',\vec x'} \right)&\equiv& {1 \over {\root 4 \of {\pi \sigma _t^2}}}\exp \left( {-{{\left( {t'-t_a} \right)^2} \over {2\sigma _t^2}}-i\omega _a\left( {t'-t_a} \right)} \right)\hfill\cr
 &&\times {1 \over {{\root 4 \of {\pi \sigma _x^2}}^3}}\exp \left( {-{{\left( {\vec x'-\vec x_a} \right)^2} \over {2\sigma _x^2}}+i\vec k_a\cdot \left( {\vec x'-\vec x_a} \right)} \right).
\label{eq.sgt.phiatime}
\end{eqnarray}
We do not require that $\omega_a=\sqrt{k_a^2+m^2}$,
although this is certainly the most obvious choice.
Given our principle of time/space symmetry
it would make sense to use the same value of the standard deviation
for both time and space,
to set $\sigma _t=\sigma _x$.
But keeping
$\sigma _t$ and $\sigma _x$
distinct will let
us compare the three and four
dimensional approaches by letting
$\sigma _t\rightarrow0$.

We will normally start with $t_a=T'$.
We define the ``lab time'' associated with a specific wave function
by
\begin{equation}
T\left[ \varphi \right]\equiv \left\langle t \right\rangle =\int {dt\varphi ^*\left( x \right)}t\varphi \left( x \right)
\label{eq.sgt.defnt}
\end{equation}
which for (\ref{eq.sgt.phiatime}) gives $T'$
as expected.
And we have for the classical position associated with a particle
\begin{equation}
\vec { {X}}\equiv \left\langle {\vec x} \right\rangle =\int {dtd\vec x\varphi ^*\left( x \right)}\vec x\varphi \left( x \right)=\vec x_a
\label{eq.piq1.avex}
\end{equation}
again as expected.

Given $\varphi$ defined at one lab time $T'$ we
get $\varphi$ at lab time $T''$
by applying the kernel (\ref{eq.piq1.freekernel})
to it
\begin{equation}
\varphi _{T''}\left( {x''} \right)=\int {dx'K_{T''-T'}\left( {x'';x'} \right)\varphi _{T'}\left( {x'} \right)}
\label{eq.piq1.phit2phit}
\end{equation}

A straightforward calculation gives
\begin{eqnarray}
\varphi _a\left( {x''} \right)&=&\varphi _a^{\left( t \right)}\left( {t''} \right)\varphi _a^{\left( {\vec x} \right)}\left( {\vec x''} \right)\hfill\cr
  \varphi _a^{\left( t \right)}\left( {t''} \right)&=&{1 \over {N_a^{{\raise3pt\hbox{$1$} \!\mathord{\left/ {\vphantom {1 8}}\right.\kern-\nulldelimiterspace}\!\lower3pt\hbox{$8$}}}}}{1 \over {\root 4 \of {\pi \sigma _t^2}}}\sqrt {{{2\pi } \over {f_t\left( {T'-T''} \right)}}}\hfill\cr
  &&\times \exp \left( {-{{\left( {t''-t_a} \right)^2} \over {2\sigma _t^2f_t\left( {T'-T''} \right)}}-i{{\omega _a\left( {t''-t_a} \right)} \over {f_t\left( {T'-T''} \right)}}-i{{\omega _a^2\left( {T'-T''} \right)} \over {2mf_t\left( {T'-T''} \right)}}} \right)\hfill\cr
  \varphi _a^{\left( {\vec x} \right)}\left( {\vec x''} \right)&=&{1 \over {N_a^{{\raise3pt\hbox{$3$} \!\mathord{\left/ {\vphantom {3 8}}\right.\kern-\nulldelimiterspace}\!\lower3pt\hbox{$8$}}}}}{1 \over {{\root 4 \of {\pi \sigma _x^2}}^3}}\sqrt {{{2\pi } \over {f_x\left( {T''-T'} \right)}}}^3\hfill\cr
  &&\times \exp \left( {-{{\left( {\vec x''-\vec x_a} \right)^2} \over {2\sigma _x^2f_x\left( {T''-T'} \right)}}+i{{\vec k_a\cdot \left( {\vec x''-\vec x_a} \right)} \over {f_x\left( {T''-T'} \right)}}-i{{\vec k_a^2\left( {T''-T'} \right)} \over {2mf_x\left( {T''-T'} \right)}}} \right)\hfill\cr
  f_{t,x}\left( T \right)&\equiv& 1+i{T \over {m\sigma _{t,x}^2}}.
\label{eq.piq1.phib}
\end{eqnarray}
Since
\begin{equation}
{{\left( {2\pi } \right)^4} \over {N_a}}=\int {dx''\,\varphi _a^*\left( {x''} \right)\varphi _a\left( {x''} \right)}
\label{eq.piq1.valueofna }
\end{equation}
we have
\begin{equation}
N_{a}=(2\pi)^{4}
\end{equation}
so the free kernels are
\begin{eqnarray}
K_{\left( {free} \right)}^{\left( t \right)}\left( {t'';t'} \right)&=&\sqrt {-{m \over {2\pi i\left( {T''-T'} \right)}}}e^{-i{m \over 2}{{\left( {t''-t'} \right)^2} \over {T''-T'}}}\hfill\cr
  K_{\left( {free} \right)}^{\left( {\vec x} \right)}\left( {\vec x'';\vec x'} \right)&=&\sqrt {{m \over {2\pi i\left( {T''-T'} \right)}}}^3e^{+i{m \over 2}{{\left( {\vec x''-\vec x'} \right)^2} \over {T''-T'}}}.
\label{eq.piq1.scaledfree4}
\end{eqnarray}
and the normalized probability distributions are
\begin{eqnarray}
p_a\left( {x''} \right)&=&p_a^{\left( t \right)}\left( {t''} \right)p_a^{\left( {\vec x} \right)}\left( {\vec x''} \right)\hfill\cr
 p_a^{\left( t \right)}\left( {t''} \right)&=&\sqrt {{1 \over {\pi \sigma _t^2\left| {f_t\left( {T'-T''} \right)} \right|^2}}}\exp \left( {-{{\left( {t''-\left( {t_a+u_0\left( {T''-T'} \right)} \right)} \right)^2} \over {\sigma _t^2\left| {f_t\left( {T'-T''} \right)} \right|^2}}} \right)\hfill\cr
 p_a^{\left( {\vec x} \right)}\left( {\vec x''} \right)&=&\sqrt {{1 \over {\pi \sigma _x^2\left| {f_x\left( {T''-T'} \right)} \right|^2}}}^3\exp \left( {-{{\left( {\vec x''-\left( {\vec x_a+\vec u\left( {T''-T'} \right)} \right)} \right)^2} \over {\sigma _x^2\left| {f_x\left( {T''-T'} \right)} \right|^2}}} \right)\hfill\cr
 u&\equiv& \left( {{{\omega _a} \over m},{{\vec k_a} \over m}} \right)
\label{eq.piq1.probb}
\end{eqnarray}
From this it is apparent
that the ``center-of-probability'' 
of the particle,
$\langle t,{\vec x} \rangle$,
is moving with four-velocity $u$.
The three velocity is $\omega/k$,
which is independent of $T$.

If we 
let $\sigma_t\rightarrow0$,
then the wave functions goes from being four to three dimensional
as does its the probability distribution.
But if we let $\sigma_t$ start finite,
then it will get still more finite (as it were) with time
$\left| {f_t\left( {T''-T'} \right)} \right|^2\sim {{T''-T'} \over {m^2\sigma ^4}}$.
In other words,
three dimensional objects stay three dimensional
and four dimensional, four.
Both assumptions are self-consistent.
It might be difficult to test either without reference to the other.

\subsection{Non-relativistic case}

Since we are primarily interested in the non-relativistic case
(we are using relativistically invariant Lagrangians
and the like only to guarantee time/space symmetry)
we ask what (\ref{eq.piq1.propagator3}) looks like in the 
non-relativistic limit.

Of course, first we need to define what we mean by the non-relativistic limit.
We take advantage of (\ref{eq.piq1.timeforce})
and define the non-relativistic limit as
being given by cases where
the integral of ${\vec v}\cdot{\vec E}$ over a typical
path is small.
In others words the acceleration 
of the time variable is small, i.e. $d\ddot {t}/dT \approx 0$.
The classical time may be approximately given
by
the linear part of (\ref{eq.piq1.tpaths})
\begin{eqnarray}
\bar {t}\left( T \right)&\approx& t'+{{t''-t'} \over {T''-T'}}\left( {T-T'} \right)\hfill\cr
 &=&t'+\gamma \left( {T-T'} \right),\ \gamma \equiv {{t''-t'} \over {T''-T'}}\hfill\cr
 \dot {\bar {t}}&\approx& \gamma 
\label{eq.piq1.timelinear}
\end{eqnarray}
and
\begin{equation}
\dot {\bar {x}}_i\approx \gamma {{d\bar {x}_i} \over {d\bar {t}}}=\gamma \bar {v}_i
\label{eq.piq1.spacelinear}
\end{equation}

The Lagrangian (\ref{eq.piq1.genl14}) becomes
\begin{eqnarray}
L&=&-{1 \over 2}m\gamma ^2+{1 \over 2}m\gamma ^2\bar {v}_i\bar {v}_i-q\gamma \phi \left( {\bar {x}} \right)+q\gamma \bar {v}_kA_k\left( {\bar {x}} \right)\hfill\cr
 &&-{1 \over 2}m\delta \dot {t}^2-{1 \over 2}q\gamma \left( {\phi _{,00}\left( {\bar {x}} \right)-\bar {v}_kA_{k,00}\left( {\bar {x}} \right)} \right)\delta t^2\hfill\cr
 &&+q\gamma \left( {\phi _{,0i}\left( {\bar {x}} \right)-\bar {v}_kA_{k,0i}\left( {\bar {x}} \right)} \right)\delta t\delta x_i\hfill\cr
 &&+{1 \over 2}m\delta \dot {x}_i\delta \dot {x}_i-{1 \over 2}q\gamma \left( {\phi _{,ij}\left( {\bar {x}} \right)-\bar {v}_kA_{k,ij}\left( {\bar {x}} \right)} \right)\delta x_i\delta x_j
\label{eq.piq1.genl15}
\end{eqnarray}
or discarding constant terms
and 
letting $\gamma\rightarrow1$
\begin{eqnarray}
L&=&{1 \over 2}m\bar {v}_i\bar {v}_i-q\phi \left( {\bar {x}} \right)+q\bar {v}_kA_k\left( {\bar {x}} \right)\hfill\cr
 &&-{1 \over 2}m\delta \dot {t}^2-{1 \over 2}q\left( {\phi _{,00}\left( {\bar {x}} \right)-\bar {v}_kA_{k,00}\left( {\bar {x}} \right)} \right)\delta t^2\hfill\cr
 &&+q\left( {\phi _{,0i}\left( {\bar {x}} \right)-\bar {v}_kA_{k,0i}\left( {\bar {x}} \right)} \right)\delta t\delta x_i\hfill\cr
 &&+{1 \over 2}m\delta \dot {x}_i\delta \dot {x}_i-{1 \over 2}q\left( {\phi _{,ij}\left( {\bar {x}} \right)-\bar {v}_kA_{k,ij}\left( {\bar {x}} \right)} \right)\delta x_i\delta x_j
\label{eq.piq1.genl16}.
\end{eqnarray}
This is the usual non-relativistic Lagrangian (as in \ref{eq.piq1.pathintegral3d})
plus terms in $\delta t^2$ and $\delta t \delta\vec x$.

This formula for the Lagrangian considerably simplifies our path integrals.
In particular, 
unless 
${\phi _{,0i}\left( {\bar {x}} \right)-\bar {v}_kA_{k,0i}\left( {\bar {x}} \right)}$
is non-zero,
the time and space parts will decouple.
If they decouple, the time part will cancel out during normalization,
just as it did in the free case
\footnote{Unless the observer at $x''$ is moving relative to the observer at $x'$.}.
Therefore,
if there is to be an effect of the $dt$ integrations,
${\phi \left( {\bar {x}} \right)-\bar {v}_kA_k\left( {\bar {x}} \right)}$ 
must depend on both time and space.
If it does not,
the four and three dimensional calculations
will give the same results 
-- at least for those cases where the semi-classical approximation is valid.

\section{A Stern-Gerlach Experiment in Time}

\subsection{Experimental arrangement}

\textit{Mutatis mutandis},
if we are to see a difference between the 
four and three dimensional calculations
we need a field which mixes space and time, e.g.
\begin{equation}
V\left( {t,\vec x} \right)\sim f\left( t \right)g\left( {\vec x} \right).
\label{eq.sgt.simple}
\end{equation}
Consider a particle with an electric dipole
moment traveling through a potential given by
\begin{equation}
\phi \left( {t,\vec x} \right)=-x\left( {E_0+E_1t} \right).
\label{eq.piq1.sgtpotential}
\end{equation}
If the electric dipole is $\vec p$,
the interaction energy is
\begin{eqnarray}
V\left( {t,\vec x} \right)&=&-\vec p\cdot \vec E\left( {t,\vec x} \right)\hfill\cr
 &=&\vec p\cdot \nabla \phi \left( {t,\vec x} \right)\hfill\cr
 &=&-p_x\left( {E_0+E_1t} \right).
\label{eq.piq1.sgtelectric}
\end{eqnarray}
\begin{figure}
\begin{center}
\includegraphics[width=3in]{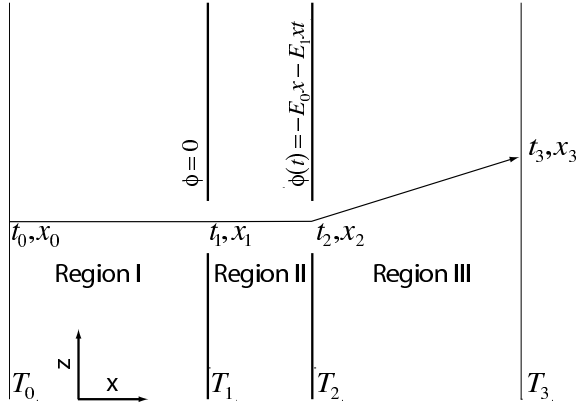}
\caption{A Stern-Gerlach experiment in time}
\label{fig:sg-time}
\end{center}
\end{figure}
Such a potential might be generated by
a capacitor perpendicular
to the $x$ axis
with the two plates
at $x_1$ and 
$x_2$ (see fig. \ref{fig:sg-time}).
A voltage applied at $x_2$
with value $-\left(E_0+E_1T\right)$,
will create a time-varying electric field
between $x_1$ and $x_2$ 
parallel to the $x$ axis
and of 
size $E_0+E_1T$.

For a quantum mechanical particle,
the value of $\vec p$ will be given by the electric dipole operator $\hat{p}$.
We will assume this has a set of eigenvalues
$\left\{ {p} \right\}$.
We will work in the basis in which
$\hat{p}$
is diagonal along the $x$ axis.
We
consider the wave function
p-component by p-component
\begin{equation}
\hat p\varphi _p\left( {x'} \right)=p\varphi _p\left( {x'} \right)
\label{eq.sgt.dipolewaves2}
\end{equation}
or
\begin{equation}
\psi \left( {x'} \right)=\sum\limits_{\left\{ p \right\}} {c_p\varphi _p\left( {x'} \right)}
\label{eq.sgt.psidipole}
\end{equation}
where the $\varphi _p\left( {x'} \right)$
are the eigenfunctions of $\hat p$.
Then
\begin{equation}
\psi \left( {x''} \right)=\sum\limits_{\left\{ p \right\}} {c_p\int {dx'\,K_p\left( {x'';x'} \right)}\varphi _p\left( {x'} \right)}.
\label{eq.sgt.psidipole2}
\end{equation}

There is a formal
resemblance between the Stern-Gerlach
experiment\cite{Gerlach:1922,Gerlach:1922b,Gerlach:1922c}
and this one.
Consider a Stern-Gerlach experiment 
with 
the beam of magnetic dipoles $\vec \mu$ going in the $+x$
direction
and with the magnetic field $\vec B=(0,0,B_{z})$
varying along the $z$ axis, $B_{z,z}\neq0$.
We may go from this
to ours
by making the replacements
\begin{eqnarray}
z&\to& t\hfill\cr
  t&\to& T\hfill\cr
  \vec \mu &\to& \vec p\hfill\cr
  B_z&\to& E_x\hfill\cr
  B_{z,z}&\to& E_{x,t}.
\label{eq.piq1.sg2sgt}
\end{eqnarray}

In most treatments of the Stern-Gerlach experiment,
the ``collapse of the wave function''
is assigned responsibility for the observed space quantization.
It would appear however
that if the finite extent of the wave function along the $z$ axis is modeled explicitly, 
e.g. as a Gaussian,
there is in fact no need to invoke the collapse;
coherent self-interference within the wave function
suffices to produce the space quantization \cite{Ashmead:2003}.
We will see a similar result here: 
coherent self-interference in time will produce time quantization.

\subsection{Electric dipole potential}

We first need to derive the correct form of the electric dipole interaction.
For two particles we have
\begin{eqnarray}
L&=&-{1 \over 2}m_1\dot {t}_1^2+{1 \over 2}m_1\dot {\vec x}_1^2-{1 \over 2}m_2\dot {t}_2^2+{1 \over 2}m_2\dot {\vec x}_2^2-V\left( {\vec x_1,\vec x_2} \right)\hfill\cr
 &&-q\dot {t}_1\phi \left( {x_1} \right)+q\dot {t}_2\phi \left( {x_2} \right)
\label{eq.sgt.L12}
\end{eqnarray}
where $V$ is the potential
that holds the dipole together.
We define
the center of mass $x$
and the relative $\tilde x$ coordinates by
\begin{eqnarray}
\matrix{{x\equiv {{m_1x_1+m_2x_2} \over {m_1+m_2}}}&{x_1=x+{{m_2} \over {m_1+m_2}}\tilde {x}}\cr
{\tilde {x}\equiv x_1-x_2}&{x_2=x-{{m_1} \over {m_1+m_2}}\tilde {x}}\cr
}
\label{eq.sgt.x2x}
\end{eqnarray}
and rewrite (\ref{eq.sgt.L12})
as
\begin{eqnarray}
L&=&-{1 \over 2}m\dot {t}^2+{1 \over 2}m\dot {\vec x}^2-{1 \over 2}\tilde {m}\dot {\tilde {t}}^2+{1 \over 2}\tilde {m}\dot {\vec {\tilde {x}}}^2-V\left( {\vec {\tilde {x}}} \right)\hfill\cr
 &&-q\left( {\dot {t}+{{m_2} \over {m_1+m_2}}\dot {\tilde {t}}} \right)\phi \left( {x+{{m_2} \over {m_1+m_2}}\tilde {x}} \right)\hfill\cr
 &&+q\left( {\dot {t}-{{m_1} \over {m_1+m_2}}\dot {\tilde {t}}} \right)\phi \left( {x-{{m_1} \over {m_1+m_2}}\tilde {x}} \right)
\label{eq.sgt.lxx}
\end{eqnarray}
with
\begin{eqnarray}
m\equiv m_1+m_2,\ \tilde {m}\equiv {{m_1m_2} \over {m_1+m_2}}
\label{eq.sgt.mm}
\end{eqnarray}

Now we assume $\tilde {x}$ small relative to ${x}$
and expand $\phi( {x}+\alpha\tilde {x})$
around $\phi( {x})$
in powers of $\alpha\tilde x$
\begin{equation}
\varphi \left( {x+\alpha \tilde x} \right)=\varphi \left( x \right)+\alpha \tilde x^\mu \partial _\mu \varphi \left( x \right)
\label{eq.sgt.phix}
\end{equation}
letting us rewrite the last two terms of (\ref{eq.sgt.lxx}) in terms of
\begin{equation}
V_{dipole}\left( {x,\tilde {x}} \right)=q\dot {\tilde {t}}\varphi \left( x \right)+q\dot {t}\tilde {t}\varphi _{,0}\left( x \right)+q\dot {t}\vec {\tilde {x}}\cdot \nabla \varphi \left( x \right)
\label{eq.sgt.3terms}
\end{equation}

The full path integral Lagrangian
is given by
\begin{equation}
{-i\varepsilon \sum\limits_{j=1}^N {\left( {{m \over 2}u_j^2+{{\tilde {m}} \over 2}\tilde {u}_j^2+V\left( {\vec {\tilde {x}}} \right)+V_{dipole}\left( {x, \tilde {x}} \right)} \right)}}
\label{eq.sgt.2fer}
\end{equation}
The $\tilde {x}$ system is not directly visible to us.
Using the cumulant approximation \cite{Khandekar:1993}
to lowest order,
we may replace the values of the relative
variables in $V_{dipole}$
by their averages
\begin{equation}
V_{dipole}\left( {x,\tilde {x}} \right)\approx q\left\langle {\dot {\tilde {t}}} \right\rangle \phi \left( x \right)+q\dot {t}\left\langle {\tilde {t}} \right\rangle \phi _{,0}\left( x \right)+q\dot {t}\left\langle {\vec {\tilde {x}}} \right\rangle \cdot \nabla \phi \left( x \right)
\label{eq.sgt.3termsave}
\end{equation}
where the average of a relative quantity $\tilde {Q}$
is defined by
\begin{equation}
\left\langle {\tilde {Q}\left( x \right)} \right\rangle =\mathop {\lim }\limits_{N\to \infty }{{\left\langle {x_N} \right|\int {\prod\limits_{i=1}^{N-1} {d\tilde {x}_i\tilde {Q}\left( {\left\{ {\tilde {x}_k} \right\}} \right)\exp \left( {-i\varepsilon \sum\limits_{j=1}^N {\left( {{{\tilde {m}} \over 2}\tilde {u}_j^2+V\left( {\vec {\tilde {x}}} \right)} \right)}} \right)\left| {x_0} \right\rangle }}} \over {\left\langle {x_N} \right|\int {\prod\limits_{i=1}^{N-1} {d\tilde {x}_i\exp \left( {-i\varepsilon \sum\limits_{j=1}^N {\left( {{{\tilde {m}} \over 2}\tilde {u}_j^2+V\left( {\vec {\tilde {x}}} \right)} \right)}} \right)\left| {x_0} \right\rangle }}}}.
\label{eq.sgt.averageq}
\end{equation}
We assume the wave functions in 
$\left| {x_0} \right\rangle$ and 
$\left| {x_N} \right\rangle$ 
are represented by some suitable time-average.
Taking (\ref{eq.sgt.3termsave}) term by term,
the first term is
\begin{equation}
\left\langle {\dot {\tilde {t}}} \right\rangle =\left\langle {\dot {\tilde {t}}_1-\dot {\tilde {t}}_2} \right\rangle ={{\left\langle {\tilde {t}''_1-\tilde {t}'_1} \right\rangle -\left\langle {\tilde {t}''_2-\tilde {t}'_2} \right\rangle } \over {T''-T'}}\approx{{\left( {T''-T'} \right)-\left( {T''-T'} \right)} \over {T''-T'}}=0
\label{eq.sgt.tilde1}
\end{equation}
where we are assuming that $\langle t''\rangle\approx T''$
and so on.

The second involves a dipole moment along the time dimension
\begin{equation}
p_0\equiv q\left\langle {\tilde {t}} \right\rangle. 
\label{eq.sgt.tilde2}
\end{equation}
To show a dipole in time,
a system has to have an asymmetry under $T$
(just as to show a permanent spatial dipole,
a system has to have an asymmetry under $P$).
We will assume our electric dipole is coming from a system symmetric under $T$,
so the second term is also zero.

We are left with the third term.
The electric dipole moment is defined as
\begin{equation}
\vec p\equiv q\left\langle {\vec {\tilde {x}}} \right\rangle 
\label{eq.sgt.tilde3}
\end{equation}
so we
have
\begin{equation}
V_{dipole}\approx-\dot {\bar {t}} {\vec p} \cdot {\vec E} \left({\bar {x}} \right)
\end{equation}
which is the same interaction as before, (\ref{eq.piq1.sgtelectric}),
times a factor of $\dot {t}$.
The contribution to the action
for a specific eigenfunction of $\hat p$
is
\begin{equation}
S_{dipole}=-\int\limits_{T_1}^{T_2} {dT\,}V_{dipole}=\int\limits_{T_1}^{T_2} {dT\,}{{dt} \over {dT}}pE\left( T \right)=p\int\limits_{t_1}^{t_2} {dt\,}E\left( t \right)=p\left\langle E \right\rangle \left( {t_2-t_1} \right)
\label{eq.sgt.actiontime}
\end{equation}

We define the impulsive approximation
as letting 
$\Delta{}T\equiv T_2-T_1\rightarrow0$ 
while
holding ${E_0\Delta T}$ and ${E_1\Delta T}$
small but finite
\begin{eqnarray}
E_0 {\Delta{}T} &\to& \textsf{E}_0\hfill\cr
 E_1 {\Delta{}T}&\to& \textsf{E}_1.
\end{eqnarray}
This approximation
corresponds to
letting the two plates of our capacitor get closer and closer
while keeping the potential across them unchanged.

\subsection{Kernel}

Since there is time but no space dependence in the interaction term,
we may focus our attention on time.
The kernel in the three space dimensions will be given by the free kernel $K^{({\vec x})}$ as in (\ref{eq.piq1.scaledfree2}).

We start with the total action in time
\begin{eqnarray}
S\left( {3;0} \right)=S_{free}\left( {3;2} \right)+S_{free}\left( {2;1} \right)+S_{free}\left( {1;0} \right)+S_{dipole}\left( {2;1} \right)
\label{eq.sgt.actiontot}
\end{eqnarray}
\begin{figure}
\begin{center}
\includegraphics[width=3in]{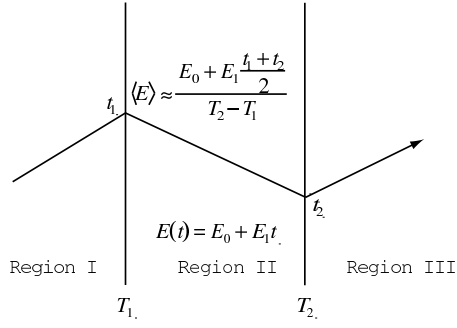}
\caption{Dipole action}
\label{fig:dipole-action}
\end{center}
\end{figure}
The interesting part is $S_{dipole}$.
To lowest order this will be given by $p\langle {E\left(t\right)} \rangle\Delta T$
where the average of $E$
is taken
over the unperturbed classical trajectory, 
a straight line
from $(T_1,t_1)$ to $(T_2,t_2)$ (see fig. \ref{fig:dipole-action}).
We define
\begin{eqnarray}
\bar T\equiv {{T_2+T_1} \over 2},\ \bar t&\equiv& {{t_2+t_1} \over 2}
\label{eq.sgt.defntt}
\end{eqnarray}
so
\begin{eqnarray}
S_{dipole}\left( {2;1} \right)\approx p\left( {E_0+E_1\bar {t}} \right)\Delta T=p\left( {\textsf{E}_0+\textsf{E}_1\bar {t}} \right)
\label{eq.sgt.actiondip4}
\end{eqnarray}

The total action $S(3;0)$ is now given by
\begin{equation}
S\left( {3;0} \right)=-{m \over 2}{{\left( {t_3-t_2} \right)^2} \over {T_3-T_2}}-{m \over 2}{{\left( {t_2-t_1} \right)^2} \over {T_2-T_1}}-{m \over 2}{{\left( {t_1-t_0} \right)^2} \over {T_1-T_0}}+pE_0+pE_1{{t_2+t_1} \over 2}
\label{eq.sgt.actiontot2}
\end{equation}
and the corresponding kernel
by
\begin{eqnarray}
K_p^{\left( t \right)}\left( {t_3,t_0} \right)&=&\sqrt {{m \over {-2\pi i \left( {T_3-T_2} \right)}}}\sqrt {{m \over {-2\pi i \left( {T_2-T_1} \right)}}}\sqrt {{m \over {-2\pi i \left( {T_1-T_0} \right)}}}\int {dt_2}dt_1\hfill\nonumber\\
 	&&\times
		\exp \left( -i{m \over 2}{{\left( {t_3-t_2} \right)^2} \over {T_3-T_2}}-i{m \over 2}{{\left( {t_2-t_1} \right)^2} \over {T_2-T_1}}-i{m \over 2}{{\left( {t_1-t_0} \right)^2} \over {T_1-T_0}} \right)
		\nonumber\\	
	&&\times	\exp \left( ip{\textsf{E}_{0}}+ip\textsf{E}_1 {{{t_2+t_1} \over 2}}  \right).
\end{eqnarray}
Rewriting $T_1$ and $T_2$ in terms of $\bar {T}$ and $\Delta T$
and discarding terms of order $\Delta T$ and higher
and of order $\textsf{E}^{2}$
we get
\begin{equation}
K_{p}^{\left( t \right)}\left( {t_3,t_0} \right)=K_{\left( {free} \right)}^{\left(t\right)}\left( {t_3;t_0} \right)
	\exp \left( ip {\textsf{E}_{0 }} \right) 
	\exp \left( {ip{\textsf{E}_1}} { { {t_3\left( {\bar {T}-T_0} \right)}+{t_0\left( {T_3-\bar {T}} \right)} } \over {T_3-T_0}} \right).
\label{eq.sgt.kernelfinal}
\end{equation}

\subsection{Wave function}

We start by assuming our initial wave function is a Gaussian in time
\begin{eqnarray}
\varphi _{T_0}\left( {t_0} \right)={1 \over {\root 4 \of {\pi \sigma _t^2}}}\exp \left( {-{{\left( {t_0-T_0} \right)^2} \over {2\sigma _t^2}}-i\omega _0\left( {t_0-T_0} \right)} \right).
\label{eq.sgt.gaussiantime}
\end{eqnarray}
We compute $\varphi(t_3)_{T_{3}}$
by applying the kernel (\ref{eq.sgt.kernelfinal})
to this
\begin{equation}
\varphi _{T_3}\left( {t_3} \right)=\int {dt_0}K_p^{\left( t \right)}\left( {t_3;t_0} \right)\varphi _{T_0}\left( {t_0} \right)
\label{ eq.sgt.phiatime2}
\end{equation}
giving
\begin{eqnarray}
\varphi _{T_3}^{\left( t \right)}\left( {t_3} \right)&=&{1 \over {\root 4 \of {\pi \sigma _t^2}}}\sqrt {{1 \over {f_t\left( {T_0-T_3} \right)}}}\hfill\cr
  &&\times \exp \left( {-{{\left( {t_3-T_0} \right)^2} \over {2\sigma _t^2f_t\left( {T_0-T_3} \right)}}-i{{\omega _3\left( {t_3-T_0} \right)} \over {f_t\left( {T_0-T_3} \right)}}-i{{\omega _3^2\left( {T_0-T_3} \right)} \over {2mf_t\left( {T_0-T_3} \right)}}} \right)\hfill\cr
  &&\times \exp \left( {ip\textsf{E}_0+ip\textsf{E}_1{{\bar T-T_0} \over {T_3-T_0}}t_3} \right)\hfill\cr
  \omega _3&\equiv& \omega _0-p\textsf{E}_1{{T_3-\bar T} \over {T_3-T_0}}\hfill\cr
  f_t\left( {T_0-T_3} \right)&=&1-i{{T_3-T_0} \over {m\sigma _t^2}}
\label{eq.sgt.wavefinal}
\end{eqnarray}
The corresponding probability distribution for a single component is given by
\begin{eqnarray}
p_{T_3}^{\left( t \right)}\left( {t_3} \right)=\sqrt {{1 \over {\pi \sigma _t^2\left| {f_t\left( {T_3-T_0} \right)} \right|^2}}}\exp \left( {-{{\left( {t_3-\left( {T_0+{{\omega _3} \over m}\left( {T_3-T_0} \right)} \right)} \right)^2} \over {\sigma _t^2\left| {f_t\left( {T_3-T_0} \right)} \right|^2}}} \right)
\label{eq.sgt.finalprob}
\end{eqnarray}
The velocity in the $x$ direction is
\begin{equation}
v_x={{\omega _3} \over {k_x}}.
\label{eq.sgt.velocity2}
\end{equation}
and
\begin{eqnarray}
{{\omega _3} \over m}\left( {T_3-T_0} \right)&=&{{\omega _o} \over m}\left( {T_3-T_0} \right)-p\textsf{E}_1\left( {T_3-\bar T} \right)\hfill\cr
  &&={{\omega _o} \over m}\left( {T_3-T_0} \right)+\Delta \omega \left( {T_3-\bar T} \right)\hfill\cr
  \Delta \omega &\equiv& -p\textsf{E}_1\hfill\cr
  &\Rightarrow& \hfill\cr
  \Delta v_x&=&{{\Delta \omega } \over {k_x}}={{-p\textsf{E}_1} \over {k_x}}.
\label{eq.sgt.velocity3}
\end{eqnarray}
The physical picture is clear.
Up to time $\bar T$ the particle is 
going with velocity $v_{x}={\omega_{0}/k_{x}}$.
At time $\bar T$ the particle gets a kick $\Delta v$
and moves with increased or decreased velocity thereafter.
When $\textsf{E}_1$ is not zero,
we get a velocity-splitting term of order $\left|p\textsf{E}_{1}\right|$
and sense given by the sign of $-\vec p\cdot\vec \textsf{E}_{1}$.

As a double-check, we ask if this is physically reasonable.
We return to the Stern-Gerlach experiment.
As noted, we may interpret the observed space
quantization as due to the extension of the wave function 
in the $z$ direction 
interacting with a magnetic field that varies in the $z$ direction.
Take, for definiteness, the case of a magnetic dipole pointing in the $+z$
direction with the
magnetic field increasing in the $+z$ direction,
$B_{z,z}>0$.
The part of the wave function on the $+z$ side of the trajectory
experiences a negative potential energy $\propto-\mu B_{z,z}$
and that on the $-z$ side a positive.
This creates an overall torque in the $+z$ direction.
To anthropomorphize slightly,
the wave function attempts to maximize its time on the 
lower energy side of the axis by turning 
towards the region with lower potential energy.

We see a similar effect here.
Take, for definiteness, the case where the electric dipole points in the $+x$
direction and the
electric field is increasing with time,
$\textsf{E}_{1}>0$.
The potential energy $\sim-p\textsf{E}_{1}$.
Anthropomorphizing again,
this particle would rather slow down,
to take advantage of the increased electric field to come.
And this is what we see in $v_{x}$;
it will be reduced by $\left|p\textsf{E}_{1}\right|$.
A particle with its electric dipole aligned in the opposite sense
 would prefer
to speed up,
to get out of the interaction region 
before the electric field gets still stronger.
It is the finite extension of the wave function in time
which opens up this kind of possibility.

For this effect to be observable,
we need the value of $\Delta \omega$
to be greater than the width of the 
wave function in energy,
$\delta\omega$.
We expect from the uncertainty principle that
$\delta\omega\sim{1/\sigma_{t}}$.
If $\sigma_{t}\to0$ then $\delta\omega\to\infty$
and the effect will be unobservable.
This is just what we said before:
$\sigma_{t}\to0$ implies the wave function is in fact three-dimensional,
so in this case no splitting in velocity should be seen.

We need an estimate of $\delta\omega$,
if we are to put the four-dimensional wave function to the test.
If we assume that the four-dimensional wave function is
composed of waves with $\omega^{2}={\vec k}^{2}+m^{2}$
then a reasonable first estimate of $\delta\omega$
is given by
\begin{eqnarray}
\delta \omega \sim {k \over \omega }\delta k
\label{eq.sgt.deltaomega2}
\end{eqnarray}
and our condition becomes
\begin{equation}
\left| {pE_1} \right|>>\left| {{k \over \omega }\delta k} \right|.
\label{eq.sgt.deltaomega3}
\end{equation}
We note analogous concerns apply to the standard Stern-Gerlach effect:
the beam must be sufficiently well localized in $k_{z}$
for the impetus $\Delta k_{z}$ from the magnetic field 
to be detectible\footnote{
There is a particularly good discussion 
of this in Baym\cite{Baym:1969}.}.

There is one other interesting term in (\ref{eq.sgt.wavefinal}),
the factor of
\begin{equation}
 \exp \left( {ip\textsf{E}_0+ip\textsf{E}_1{{\bar T-T_0} \over {T_3-T_0}}t_3} \right).
\end{equation}
(This does not contribute to the probability distribution
because it is purely oscillatory.)
If we hold $T_{0}$ and $\bar T$ fixed,
while letting $T_{3}\to\infty$,
we get a change in frequency of
\begin{equation}
-\left(p\textsf{E}_{0}+p\textsf{E}_{1}\bar T\right)
\label{eq.sgt.precession}
\end{equation}
which is just the precession predicted by the \seqn (see (\ref{eq.sgt.seqndeltafreq}) below).

\subsection{Results using \seqn}

To complete the analysis we now solve the same problem
using the \seqn.
\begin{eqnarray}
i{\partial \over {\partial T}}\psi \left( {T,\vec x} \right)=-{1 \over {2m}}\nabla ^2\psi \left( {T,\vec x} \right)-\vec p\cdot \vec E\psi \left( {T,\vec x} \right)
\label{eq.piq1.sgtseqn}
\end{eqnarray}
or if we write $\psi$ in terms of the $p$ components of the $\hat{p}$ operator
\begin{equation}
i{\partial  \over {\partial T}}\varphi _p\left( {T,\vec x} \right)=-{1 \over {2m}}\nabla ^2\varphi _p\left( {T,\vec x} \right)-p\left( {E_0+E_1T} \right)\varphi _p\left( {T,\vec x} \right)
\label{eq.sgt.seqnpx}
\end{equation}

We may solve using separation of variables.
\begin{equation}
\psi _p\left( {T,\vec x} \right)=\xi _p\left( T \right)\chi \left( {\vec x} \right)
\label{eq.sgt.wfsep}
\end{equation}
giving
\begin{eqnarray}
\left( {i{\partial \over {\partial T}}+{{\vec k^2} \over {2m}}} \right)\xi_p \left( T \right)&=&-p\left( {E_0+E_1T} \right)\xi_p \left( T \right)\hfill\cr
 {{\vec k^2} \over {2m}}\chi \left( {\vec x} \right)&=&-{1 \over {2m}}\nabla ^2\chi \left( {\vec x} \right).
\label{eq.sgt.seqnsep}
\end{eqnarray}

To lowest order,
the change in frequency per component 
from the dipole interaction
will be given by
\begin{eqnarray}
\Delta \omega ={1 \over {\Delta T}}\int\limits_{T_1}^{T_2} {dT'\,\left( {-p\left( {E_0+E_1T'} \right)} \right)}=-p\textsf{E}\left( {\bar T} \right)
\label{eq.sgt.seqndeltafreq}
\end{eqnarray}
which is the same as the earlier result(\ref{eq.sgt.precession}). 
We see no $\Delta v$ term
and therefore no splitting of the beam in time.

This implies that
a time sensitive detector will see one hump if the \seqn\ 
is correct,
but two (or more) humps if four dimensional path integrals should be used.
If the eigenfunctions of the electric dipole are $\left\{ p_{i} \right\}$
then the humps will be spaced in velocity by $-\left(p_{i+1}-p_{i}\right)\textsf{E}_{1}$,
and if this spacing is greater than $\delta \omega \sim ({\omega/k}) \delta k$,
they should be observable.
 
\subsection{Comparison of four dimensional path integrals to \seqn}

So there is a clear difference between the
results with
four dimensional path integrals and the \seqn.
With the \seqn\ we predict the electric dipole will precess around the $x$ axis,
but we do \emph{not} predict a split
of the beam
in velocity or time.
With four dimensional path integrals
we predict the change in precession and 
\emph{in addition} that the beam will be split
in velocity and time.

This is a non-relativistic effect;
we used a Lorentz invariant Lagrangian
solely to ensure time/space symmetry.

The splitting in velocity is not induced by the collapse of the wave function;
we ``used up'' the collapse when we broke the incoming wave function up into
eigenfunctions of the electric dipole operator.

The proposed experiment is clearly only a very crude test 
of these ideas.
A subtler approach would be required to see the effect of the
second order corrections to the action.

The principle advantage of the approach given here 
is the much greater symmetry between time and space. 
The principle disadvantage is that it is not entirely clear what is meant by 
this greater symmetry between time and space.
The Machian approach to time suggested above may 
provide a useful line of attack.

\section{Discussion}

We have observed that the assumptions about time
implicit in the \seqn\ 
and in path integrals
are very different.
The \seqn\ takes a classical view of time,
seeing time instant by instant
and 
using the present to define the future.
Path integrals most naturally see time all at once,
from the `block universe perspective'.
This suggests that the two formalisms 
may not be completely equivalent.
And that therefore it could be useful to quantize
path integrals independently,
working from first principles.

We have done this by
starting with an abstract formula for path integrals
and
imposing two requirements:
correct 
behavior in the classical limit
and
the most complete practicable symmetry between time and space.
We refer to this as ``path integral quantization''
to emphasize that we are taking path integrals as our starting point.

Path integral quantization predicts
quantum fluctuations over the time dimension
analogous to the quantum fluctuations seen over the three space dimensions.
For constant potentials 
there is no effect.
But in the presence of rapidly varying electromagnetic fields,
the coupling between these quantum fluctuations in time
and the fields
should be detectable.

We considered in particular the case
where a particle with a non-zero electric dipole moment
is sent along the $x$ axis though a rapidly varying
electric field, also along the $x$ axis.
The
\seqn\ predicts precession around 
the $x$ axis
but no physical splitting of the beam.
Path integral quantization
predicts the precession,
\emph{and} that the beam
will be split in velocity and time.

Path integral quantization may be generalized to include QED:
sums over $\vec k$ become sums over $\omega $ and $\vec k$,
the condition $\omega^2=\vec k^2+m^2$ 
becomes $\omega^2\approx\vec k^2+m^2$
(quantum fluctuations of $\omega$ around $\sqrt{\vec k^2+m^2}$
are permitted),
and
normalization and renormalization have to be handled from the 
block universe perspective.
But this is beyond the scope of this work.

\begin{acknowledgments}
\textsc{Acknowledgments}

\rm I would like to thank Jonathan Smith 
	for invaluable encouragement, guidance, and practical assistance.

I would like to thank 
	Catherine Asaro, 
	John Cramer, 
	the late Robert Forward, 
	Dave Kratz,
	and Joseph Sedlak
	for helpful conversations.

I would like to thank my friends 
	Gaylord Ashmead, 
	Graham Ashmead, 
	Diane Dugan, 
	Linda Kalb, 
	Steve Robinson, and 
	Ferne Welch 
	for encouragement and support.

I would like to thank the librarians of 
	Bryn Mawr College, 
	Haverford College, and the 
	University of Pennsylvania 
	for their unflagging helpfulness.
	
And I would like to express my appreciation 
	of the great work done by the 
	administrators of the invaluable electronic archives at 
	PROLA, LANL, and SLAC.
	
\end{acknowledgments}

\bibliography{piq}

\end{document}